\documentclass{elsarticle}

 \usepackage{graphicx}

\usepackage{amssymb,dsfont}
\usepackage{latexsym}
\input{epsf}

\newcommand{\simgt}{\lower.5ex\hbox{$\; \buildrel > \over \sim \;$}}
\newcommand{\simlt}{\lower.5ex\hbox{$\; \buildrel < \over \sim \;$}}

\newcommand{\be}{\begin{equation}}
\newcommand{\ee}{\end{equation}}
\newcommand{\bea}{\begin{eqnarray}}
\newcommand{\eea}{\end{eqnarray}}
\newcommand{\bi}{\begin{itemize}}
\newcommand{\ei}{\end{itemize}}

\newcommand{\bfi}{\begin{figure}
\epsfxsize=9cm 
\epsffile}
\newcommand{\efi}{\end{figure}}
\newcommand{\la}{\lesssim}
\def\gsim{ \lower .75ex \hbox{$\sim$} \llap{\raise .27ex \hbox{$>$}} }
\def\lsim{ \lower .75ex \hbox{$\sim$} \llap{\raise .27ex \hbox{$<$}} }

\begin{document}

\begin{frontmatter}
\title{Cosmological Tests of Gravity }
\author{Bhuvnesh Jain \& Justin Khoury}
\address{Center for Particle Cosmology, Department of Physics and Astronomy,\\ University of Pennsylvania,
  Philadelphia, PA 19104}
\ead{bjain@physics.upenn.edu, jkhoury@physics.upenn.edu}






\begin{abstract}

Modifications of general relativity provide an alternative explanation
to dark energy for the observed acceleration of the universe.  
We review recent developments in modified gravity theories, focusing
on higher dimensional approaches and chameleon/$f(R)$ theories. We classify
these models in terms of the screening mechanisms that enable such theories 
to approach general relativity on small scales (and thus satisfy solar system constraints).
We describe general features of the modified Friedman equation in such
theories. 

The second half of this review describes experimental tests of gravity in light of the new
theoretical approaches. We summarize  
the high precision tests of gravity on laboratory and solar system scales. 
We describe in  some detail  tests on astrophysical scales ranging from  
$\sim$ kpc (galaxy scales) to $\sim$ Gpc (large-scale structure).  These tests  
rely on the growth and inter-relationship of perturbations 
in the metric potentials, density and  velocity fields
which can be measured using gravitational lensing,  galaxy cluster abundances, galaxy
clustering and the Integrated Sachs-Wolfe effect. A robust way to interpret
observations is by constraining effective parameters, such as the
ratio of the two metric potentials. Currently tests of gravity  on astrophysical 
scales are in the early stages --- we summarize these tests and 
discuss the interesting prospects for new tests in the coming decade. 
\end{abstract}

\end{frontmatter}
\begin{tableofcontents} 
\end{tableofcontents}

\section{Introduction}
\label{intro}

Einstein's theory of general relativity (GR) has proven spectacularly successful over 90 years 
of experimental tests~\cite{Will2001}. These tests range from millimeter scale tests in the laboratory to solar
system tests and consistency with gravity wave emission by binary pulsars. Recently cosmological motivations
for modifying gravity have renewed interest in testing GR on scales of galaxies and large-scale structure. 

Over the last decade it has become clear that the energy contents of the universe pose a 
major puzzle, in that GR plus the Standard Model of particle
physics can only account for about $4\%$ of the energy density inferred from
observations. By introducing dark matter and dark  
energy, which account for the remaining $96\%$ of the total 
energy budget of the universe, cosmologists have been able to
account for a wide range of observations, from the overall expansion of the universe to 
the large scale structure of the early and late universe~\cite{Reviews}. 

The $\Lambda$-Cold Dark Matter ($\Lambda$CDM) paradigm has thus emerged as the standard model of
cosmology. While its phenomenological success is undeniable, few would argue for its aesthetic appeal.
The model requires introducing exotic components of matter and energy. And while the vacuum energy of quantum fields
offers a natural candidate for dark energy, the predicted amplitude from field theory calculations is many orders of magnitude
larger than inferred from observations~\cite{weinberg}. 

A critical assumption in the standard picture is the validity of GR at galactic and cosmological scales.
Since GR has not been tested independently on these scales, a natural alternative is that there are new gravitational
degrees of freedom associated with the dark energy scale, and that these degrees of freedom are responsible for the observed late-time
cosmic acceleration. This possibility, that modifications in GR at cosmological scales can replace dark matter and/or
dark energy, has become an area of active research in recent years. See~\cite{markalessandra} and references therein. The driving motivation for many practitioners is of course the
vexing problem of the cosmological constant, since any solution to this long-standing puzzle promises to have profound implications for the nature of cosmic acceleration. 

This review article discusses the prospects of testing GR on the largest scales through cosmological
observations. Doing so in a model-independent way is unfortunately not without ambiguities. Unlike solar system tests, where the matter stress-energy
tensor is known ({\it i.e.}, the Sun), cosmological tests inherently rely on assumptions about the stress-energy composition
of the universe --- any modification to the gravitational side of Einstein's equations can equivalently be interpreted
as a dark energy contribution to the stress-energy side~\cite{kunz,husawicki,costaseff}. Throughout this paper, we assume that the dark energy
component is nearly spatially-homogeneous. Thus strictly speaking the cosmological tests discussed
below are in fact tests of the assumption of GR plus smooth dark energy. Fortunately, however, nearly all of the specific
modifications discussed below also lead to testable deviations from GR in the solar system.

In the rest of this introductory section, we briefly review existing constraints on deviations from GR from solar system and laboratory tests of gravity.
In Sec.~\ref{MGT} we review some of the leading proposals for modified gravity (MG) theories, designed as an alternative to dark energy to explain the present day 
acceleration of the universe. In these models gravity at late cosmic times and on large scales departs from the predictions of GR. Note that for the purpose of this review
we do not discuss alternatives to dark matter, such as Modified Newtonian Dynamics (MOND)~\cite{MOND} and its Tensor-Vector-Scalar (TeVeS) covariant generalization~\cite{teves,costasteves}.
In Sec.~\ref{tests} we consider the prospects of testing such models in view of the massive new astronomical surveys and other experiments expected in the coming decade.
We conclude with brief remarks in  Sec.~\ref{discuss}.

\subsection{Approaches to Modify Gravity} 

The field equations of GR can be derived from the action
\be
S_{\rm GR} = \frac{M_{\rm Pl}^2}{2}\int {\rm d}^4x\sqrt{-g}\ R + S_{\rm matter}[g_{\mu\nu}]\,, 
\label{GR_Action}
\ee
where $M_{\rm Pl}^2=1/8\pi G_{\rm N}$ with $G_{\rm N}$ being Newton's gravitational constant, $g$ is the determinant of the metric tensor $g_{\mu\nu}$, and $R$ is the Ricci scalar. The first term is the Einstein-Hilbert action, while $S_{\rm matter}$ contains all matter fields, with minimal couplings to $g_{\mu\nu}$. The stationary point of the variation of $S_{\rm GR} $ with respect to the metric then yields the Einstein field equations. 

The simplest and best-studied modifications to General Relativity are scalar-tensor theories, which introduce a scalar cousin
to the graviton. First proposed by Brans and Dicke~\cite{BD} as an attempt to reconcile gravity with
Machian ideas~\cite{Mach,bucket,wheeler,khouryparikh}, scalar-tensor lagrangians are now understood to arise as low-energy limits of various theories of particle physics.
These include dimensional reductions of higher-dimensional gravity and brane-world models~\cite{ADD,hetMtheory,RS,cedricbrane,branereview,renjie}. Moreover, a host of more intricate modifications of gravity, including massive gravity and the Dvali-Gabadadze-Porrati (DGP) model~\cite{DGP,DGP2,DGP3,luerev}, reduce to a scalar-tensor form in certain decoupling limits~\cite{ags,luty,nicolis}.

The Brans-Dicke action~\cite{BD} is characterized by a universally-coupled scalar field,
\be
S_{\rm BD} = \frac{M_{\rm Pl}^2}{2}\int {\rm d}^4x\sqrt{-\tilde{g}}\left(\Phi \tilde{R} - \frac{\omega_{\rm BD}}{\Phi}(\partial\Phi)^2\right) +
S_{\rm matter}[\tilde{g}_{\mu\nu}]\,.
\label{BDjordan}
\ee
The effective Newton's constant is thereby promoted to a scalar field. The field redefinitions $g_{\mu\nu} = \Phi \tilde{g}_{\mu\nu}$ and $\phi = -M_{\rm Pl}\sqrt{3/2 + \omega_{\rm BD}} \log \Phi$ map~(\ref{BDjordan}) to the Einstein frame, where the coefficient of the Einstein-Hilbert term is constant:
\be
S_{\rm BD}^{\rm Einstein} = \int {\rm d}^4x\sqrt{-g}\left(\frac{M_{\rm Pl}^2}{2} R - \frac{1}{2}(\partial\phi)^2\right) + S_{\rm matter}\left[g_{\mu\nu} e^{2\beta\phi/M_{\rm Pl}}\right] \,.
\label{BDeinstein}
\ee
For future use, we have introduced $2 \beta \equiv 1/\sqrt{3/2 + \omega_{\rm BD}}$. Unfortunately, as reviewed below, the Brans-Dicke parameter is so tightly constrained by solar system and pulsar observations,
$\omega_{\rm BD} \;\gsim\; 4\times 10^4$~\cite{Will2001}, that the cosmological effects of the scalar field are rendered uninterestingly small. Since this stringent constraint only applies to visible matter, many authors have relaxed the assumption of universal coupling and explored models in which the scalar field only substantially interacts with the dark matter~\cite{dmde,amendola,wei,
peebles04,hueywandelt,CarrollTrodden,piersteDMDE,ravi1,ravi2}.

The alternative gravity theories discussed in Sec.~\ref{MGT} all reduce to scalar-tensor theories in certain limits. (See~\cite{piazza}, however, for an IR modification of gravity
without new degrees of freedom.) To produce interesting effects cosmologically while satisfying the tight local constraints on deviations from GR, these scalar fields must somehow be hidden or screened in the local environment. There are three known mechanisms for screening a scalar field, all of which exploit the fact that the matter density at a typical location in the solar system or in pulsar environments
is many orders of magnitude larger than the mean cosmic density. The success of these models relies on a nonlinear mechanism that operates on
small scales or inside dense objects to recover GR on solar-system/pulsar scales where gravity is well tested. The large local background density
triggers non-linearities in the scalar field, which in turn result in its decoupling from matter. In short, the three mechanisms rely on giving the scalar field
a large mass, a large inertia, or by weakening its coupling to matter, respectively.

The first mechanism, discussed in Sec.~\ref{cham}, arises in chameleon field theories~\cite{cham1,cham2,cham3,shawpapers}. By adding a suitable self-interaction potential $V(\phi)$ to~(\ref{BDeinstein}), the chameleon scalar field acquires mass which depends on the density. The mass is large in regions of high density, thereby suppressing any long-range interactions. (Density-dependent effective couplings were initially noted in a different context~\cite{damourespositofarese}.) Chameleon fields can in fact have different couplings to different matter species, thereby generalizing the universal coupling assumed in Brans-Dicke theories. A subclass of chameleon theories are so-called $f(R)$ theories of gravity~\cite{fR,f(R)2,f(R)3,f(R)rev,f(R)review,HuSa,Staro} --- for $\omega_{\rm BD} = 0$ and suitable choice of $V(\phi)$, the scalar-tensor action can be mapped through field redefinitions into an action of the form~(\ref{GR_Action}) with the Einstein-Hilbert term replaced by $f(R)$, a general function of the Ricci scalar.

The second mechanism for hiding a scalar is achieved with symmetron fields (Sec.~\ref{symm}), proposed recently in~\cite{symmetron,symmetronearlier}. The symmetron Lagrangian is qualitatively similar to that of chameleon models, but the mechanism and its phenomenological consequences are drastically different. The screening in this case relies on a scalar field acquiring a vacuum expectation value (VEV) that is small in high-density regions and large in low-density regions. An essential ingredient is that the coupling to matter is proportional to this VEV, so that the scalar couples with gravitational strength in low-density environments, but is decoupled and screened in regions of high density. This is achieved through a symmetry-breaking potential, hence the name symmetron.

The third mechanism, discussed in Sec.~\ref{vain}, relies on the
scalar field having derivative interactions that become large in
regions of high density or in the vicinity of massive
objects. Perturbations of the scalar in such regions acquire a large
kinetic term and therefore decouple from matter. Thus the scalar
screens itself and becomes invisible to experiments. This so-called
Vainshtein effect~\cite{vainshtein,ddgv,ziour} ensures the phenomenological
viability of brane-world modifications of
gravity~\cite{DGP,DGP2,DGP3,DGP4,DGP5,cascade1,cascade2,nonFP,cascade3,cascade4,aux,claudiaaux,nishant}
and galileon scalar field theories~\cite{galileon,deff,nathan,kmouf,galileoncosmo1,galileoncosmo2,galileoncosmo3}. 

These various theories and the corresponding observational constraints will be discussed respectively in Secs.~\ref{MGT} and~\ref{tests}.
To set the stage for these discussions, we briefly review the current constraints on GR from laboratory and solar system tests of gravity.
For a more detailed review of the subject, see~\cite{Will2001}.

\subsection{Newtonian Tests}
\label{newtonian}

Einstein's gravity theory is based on the weak equivalence principle (WEP), which states that the trajectory of a 
freely falling test body is independent of its internal structure and composition. The WEP is of course not unique to GR ---
any theory whose matter fields couple to a unique metric tensor, {\it e.g.} the Brans-Dicke theory in~(\ref{BDjordan}), 
satisfies the WEP, independent of the field equations governing this metric. (As we review below, scalar-tensor theories, including Brans-Dicke,
violate a stronger version of the equivalence principle satisfied in GR --- the Strong Equivalence Principle (SEP) --- which extends the
universality of free-fall to include gravitational self-energy contributions.)

E$\ddot{{\rm o}}$tvos-type experiments test the WEP by measuring the $\eta$ parameter --- the fractional difference in the acceleration
of freely falling bodies of different composition. The best limit comes from the torsion balance experiment of the University of Washington
(E$\ddot{{\rm o}}$t-Wash), which gives $\eta = (0.3\pm 1.8) \times 10^{-13}$~\cite{etaparam}. The forthcoming space-based mission
MicroSCOPE~\cite{microscope} is designed to reach $10^{-15}$ sensitivity. Other satellite experiments, such as Galileo Galilei~\cite{GG}
and the Satellite test of the Equivalence Principle (STEP)~\cite{STEP}, will further improve on current limits.

Even when the WEP is satisfied, modifications of gravity are also constrained by tests of the gravitational inverse-square law~\cite{fischbach}. The simplest example
is a scalar field with mass $m$ mediating a force of range $\lambda$ and coupling strength $\alpha$, corresponding to the Yukawa potential
\begin{equation}
\psi = -\alpha\frac{G_{\rm N}M}{r} e^{-r/\lambda}
\label{eqn:Yukawa}
\end{equation}
Experimental tests can be characterized as providing limits on $\lambda$ and $\alpha$~\cite{etaparam,adel,eotwash}. For gravitational-strength coupling ($\alpha\sim {\cal O}(1)$),
there is no evidence of a fifth force down to $\lambda = 56\;\mu$m~\cite{adel}.

\subsection{Post-Newtonian Tests}
\label{postnewtonian}

The early experimental successes of GR were the consistency with the
perihelion shift of Mercury and the deflection of light by the
Sun. Beginning around 1960, a number of experimental tests of GR were
carried out, including the measurement of gravitational redshift. The
consistency of the decrease in the orbital period of the Hulse-Taylor
binary pulsar with the GR prediction of energy loss due to gravity
waves further established the validity of GR in the weak field
regime. 

Solar system tests of alternate theories of gravity are commonly phrased in the language of the 
Parameterized Post-Newtonian (PPN) formalism. The PPN formalism applies to metric
theories of gravity in the weak field, slow motion limit, {\it i.e.} when potentials and 
velocities are small: $\Psi, v^2/c^2 \sim \epsilon^2 \ll 1$. The metric can then be written 
as a perturbation about the Minkowski metric (or, as we shall see, the FRW metric for the
expanding universe). Terms up to second order in the potentials are included, with coefficients
that are unity or zero for GR replaced by parameters that accommodate MG theories. 

The $\gamma$ and $\beta$ PPN parameters are given by the metric
\begin{equation}
ds^2 = -(1 + 2\Psi - 2\beta\Psi^2)\ {\rm d}t^2 + (1 - 2\gamma\Psi)\ {\rm d}\vec{x}^2
\label{eqn:PPN}
\end{equation}
where the potential $\Psi=-G_{\rm N}M/r$ for the Schwarzschild metric. The parameter $\gamma$ describes the spacetime curvature induced by a unit mass and $\beta$ the nonlinearity in the superposition law of gravity. We will employ a similar metric in Sec.~3 with the FRW metric and an arbitrary (small) potential. In general for a fluid description of matter, allowing for generic Poisson-like potentials, the PPN formalism requires ten parameters for a complete description~\cite{Will2001}. Note that the PPN formalism has constant parameters and therefore does not accommodate Yukawa-like modifications with finite $\lambda$. For astrophysical tests we will work with effective parameters that may have a scale and time dependence, but a similar expansion of the metric is still useful. 

The Brans-Dicke theory given by~(\ref{BDjordan}) has identical PPN parameter values as in GR, except for $\gamma = (1+\omega_{\rm BD})/(2+\omega_{\rm BD})$.
Since $\gamma$ is the most relevant PPN parameter for the modified gravity theories of interest, we focus on this parameter for the rest of our discussion.

The tightest constrain on $\gamma$ comes from time-delay measurements in the solar system, specifically the Doppler tracking of the Cassini spacecraft,
which gives $\gamma - 1 = (2.1\pm 2.3)\times 10^{-5}$~\cite{bertotti}. Light deflection measurements, meanwhile, constrain
$\gamma$ at the $10^{-4}$ level~\cite{shapiro}. The observed
perihelion shift of Mercury's orbit sets a weaker limit of $10^{-3}$~\cite{mercury}. 

As mentioned earlier, GR satisfies an extended version of the equivalence principle, the SEP, which states, in particular, that macroscopic objects follow the same trajectory in 
 a uniform gravitational field as test masses. In other words, the universality of free-fall is preserved in GR even when accounting for self-gravity contributions. The SEP is violated
 in Brans-Dicke theories and in all of the MG theories considered here. Violations of the SEP result in the Nordtvedt effect~\cite{nordtvedt} --- a difference in the free-fall acceleration of the Earth and the Moon towards the Sun, which is detectable in Lunar Laser Ranging (LLR). (Because the Earth and the Moon have different compositions, however, one must worry about a fluke cancellation between WEP and SEP violations in this measurement. To disentangle these effects, laboratory tests of the WEP have been performed using tests masses with Earth-like and Moon-like compositions~\cite{baessler}.) Searches for the Nordtvedt effect in LLR data constrain PPN deviations from GR at the $10^{-4}$ level~\cite{Will2001}.

\section{Modified Gravity Theories}
\label{MGT}

In this Section we review various proposals for modifying GR at large distances. The list is by no means exhaustive --- our goal is to provide the reader
with an overview of a few broad classes of theories that have attracted considerable interest over the last few years, highlighting key theoretical and
observational differences among them. For this purpose, we find it useful to group theories based on the three qualitatively different non-linear mechanisms through
which GR is approximately recovered locally. These are the chameleon (Sec.~\ref{cham}), Vainshtein (Sec.~\ref{vain}) and symmetron (Sec.~\ref{symm}) mechanisms.

\begin{figure}[h!] 
\begin{center}
\includegraphics[width=0.8\textwidth]{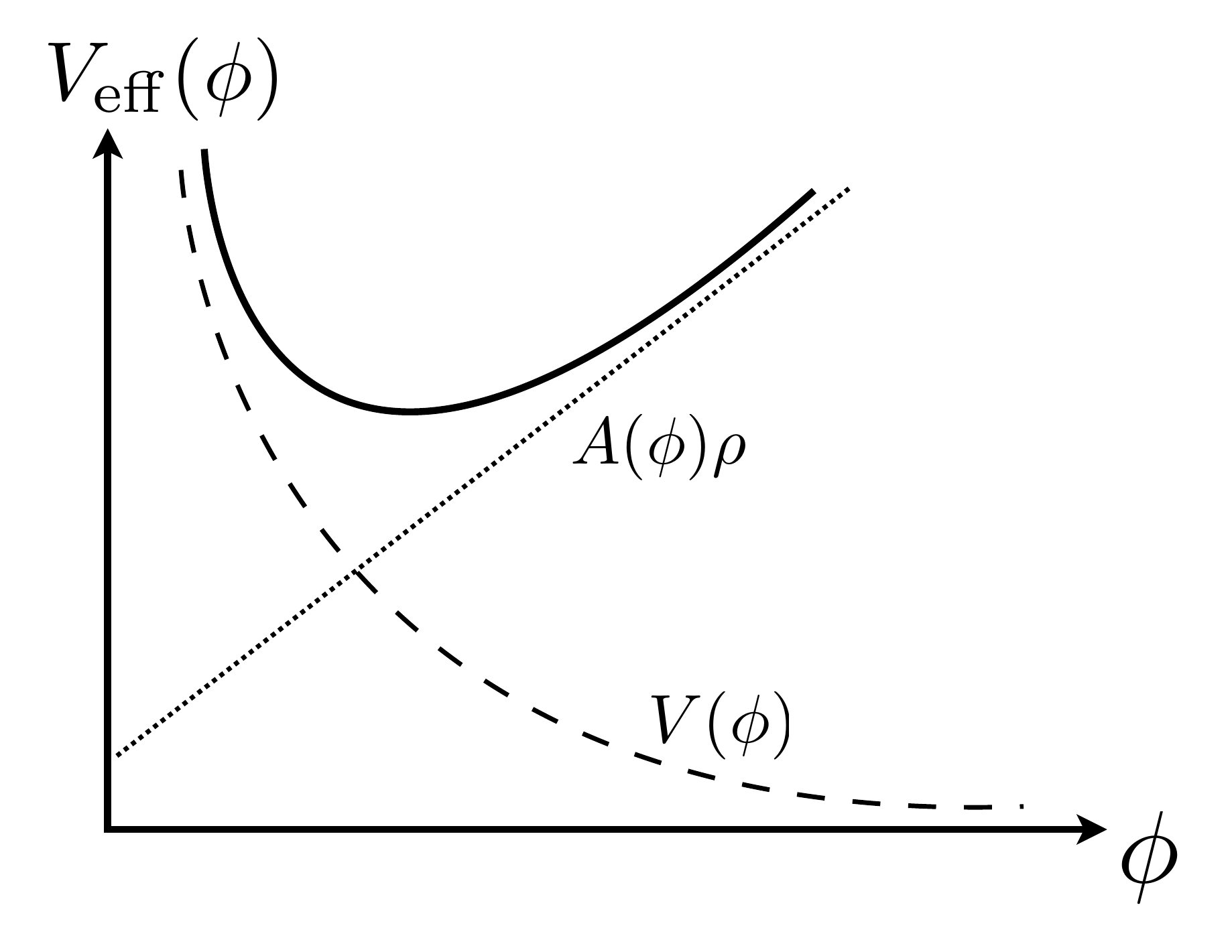}
\end{center}
\caption{The chameleon effective potential $V_{\rm eff}$ (solid curve) is the sum of two contributions: the actual potential $V(\phi)$ (dashed curve), plus a density-dependent term from its coupling to matter (dotted curve).}
\label{poteff}
\end{figure}

\subsection{Chameleon/$f(R)$ Field Theories}
\label{cham}

Chameleon field theories generalize~(\ref{BDeinstein}) to include a scalar potential $V(\phi)$, whose properties will be discussed shortly, as well as a
more general coupling $A(\phi)$ to matter fields:
\be
S_{\rm cham} = \int {\rm d}^4x\sqrt{-g}\left(\frac{M_{\rm Pl}^2}{2} R - \frac{1}{2}(\partial\phi)^2 - V(\phi)\right) + S_{\rm matter}\left[g_{\mu\nu}A^2(\phi) \right] \,.
\label{Scham}
\ee
One can allow different couplings to the various matter fields, thereby introducing violations of the Equivalence Principle. Furthermore, one can also
introduce a coupling to the electromagnetic field strength, which induces photon-chameleon mixing in the presence of magnetic fields~\cite{brax2,burrage}.
For the purpose of this review article, we focus on the simplest case of a universal, conformal coupling of the chameleon, as in~(\ref{Scham}).

The equation of motion for $\phi$ that derives from this action is
\be
\Box\phi = V_{,\phi} - A^3(\phi)A_{,\phi} \tilde{T}\,,
\label{phigen}
\ee
where $\tilde{T} = \tilde{g}_{\mu\nu} \tilde{T}^{\mu\nu}$ is the trace of the energy-momentum tensor defined with respect to the
Jordan-frame metric $\tilde{g}_{\mu\nu} = A^2(\phi) g_{\mu\nu}$. Since matter fields couple minimally to $\tilde{g}_{\mu\nu}$, this stress tensor is covariantly
conserved: $\tilde{\nabla}_\mu  \tilde{T}^{\mu\nu} = 0$.

To study the field profile on solar system and galactic scales, we can approximate the metric in~(\ref{phigen}) as flat space, ignore time derivatives, and focus on 
the case of a non-relativistic, pressureless source. In terms of an energy density $\rho = A^3(\phi)\tilde{\rho}$ conserved in Einstein frame, we obtain
\be
\nabla^2\phi = V_{,\phi} + A_{,\phi}\rho\,.
\label{phistat}
\ee
Thus, because of its coupling to matter fields, the scalar field is affected by ambient matter density. Its profile is governed by an effective potential
\be
V_{\rm eff}(\phi) = V(\phi) +A(\phi)\rho \,.
\label{Veffcham}
\ee

For suitably chosen $V(\phi)$ and $A(\phi)$, this effective potential can develop a minimum  at some finite field value $\phi_{\rm min}$ in the presence of background matter density, as illustrated in Fig.~\ref{poteff}, where the mass of the chameleon field is sufficiently large to evade local constraints:
\be
m_{\rm min}^2 = V_{,\phi\phi}(\phi_{\rm min}) + A_{,\phi\phi}(\phi_{\rm min})\rho\,.
\ee
Assuming $A(\phi)$ is monotonically increasing, for concreteness, the general conditions that $V(\phi)$ must satisfy are~\cite{cham1,brax1}: (i) to balance the potential against the density term, we must have $V_{,\phi} < 0$ over the relevant field range; (ii) since $V_{,\phi\phi}$ typically gives the dominant contribution to the mass term, stability requires $V_{,\phi\phi}> 0$; (iii) the effective mass increases with density provided that $V_{,\phi\phi\phi} < 0$.

As seen from Fig.~\ref{poteff}, for suitable choices of $V$ and $A$, the effective mass can be a steeply growing function of the ambient density.
The chameleon mechanism therefore exploits the large difference between local ($\sim 1~{\rm g}/{\rm cm}^3$) and cosmic ($\approx 10^{-30}~{\rm g}/{\rm cm}^3$)
densities to generate a wide range in mass scales. As we have seen in Sec.~\ref{newtonian}, tests of the inverse-square law constrain the range in the laboratory
to be $\lsim\; 50~\mu$m. Meanwhile, we would ideally like to achieve $m_{\rm eff}\sim H_0$ at cosmological densities, such that the chameleon behaves as a
quintessence field. In all known examples, however, the best one can do is $m_{\rm eff} \;\gsim \; 10^4\;H_0$, corresponding to a range $\lsim \; 0.1$~Mpc in the cosmos. 
The reason for this is simple --- for a potential for which $m_{\rm eff}\sim H_0$ at cosmological densities, the field value where this is reached is generally a
distance $\sim M_{\rm Pl}$ away from the local value, and maintaining gradient of such a large field difference is just energetically too costly. The largest gradient energy
that can be tolerated corresponds in practice to $m_{\rm eff} \;\gsim \; 10^4\;H_0$ at cosmological densities.

A prototypical potential that satisfies all of the above conditions is the inverse power-law form, 
\be
V(\phi) = \frac{M^{4+n}}{\phi^n}\,,
\label{inversepot}
\ee
where $n$ is some positive constant. This falls within the class of tracker potentials relevant for quintessence models of dark energy~\cite{zlatev}. 
For the coupling function, a nice choice that makes contact with Brans-Dicke theories is the exponential form of~(\ref{BDeinstein}):
\be
A(\phi) = e^{\beta\phi/M_{\rm Pl}}\,.
\label{coup}
\ee
The parameter $\beta$ is implicitly assumed to be~${\cal O}(1)$, corresponding to gravitational strength coupling. Remarkably, it was pointed out in~\cite{shawpapers} that
much larger couplings are allowed by current constraints, but one must be concerned with adiabatic instabilities for $\beta\gg 1$~\cite{markandrachel}.
Assuming $\beta\phi\ll M_{\rm Pl}$, which will be the case for most situations of interest, the effective minimum lies at
\be
\phi_{\rm min} \approx \left(\frac{nM^{4+n}M_{\rm Pl}}{\beta\rho}\right)^{\frac{1}{n+1}}\,.
\label{phimin}
\ee
Meanwhile, the mass of small fluctuations around this value is
\be
m_{\rm min}^2 \approx n(n+1) M^{-\frac{4+n}{1+n}}\left(\frac{\beta\rho}{nM_{\rm Pl}}\right)^{\frac{n+2}{n+1}}\,.
\ee
Both $\phi_{\rm min}$ and $m_{\rm min}$ manifestly depend on the matter density --- $m_{\rm min}$ grows larger for increasing $\rho$, as desired, while $\phi_{\rm min}$ decreases.

Potentials with {\it positive} powers of the field, $V(\phi) \sim \phi^{2\alpha}$ with $\alpha$ an integer $\geq 2$, are also good candidates for chameleon theories.
In the simplest case of $V(\phi) = \lambda \phi^4$,~\cite{cham2} showed that existing laboratory constraints at the time were satisfied for $\beta = 1$ and $\lambda = 1$. Moreover, the authors of~\cite{cham2} identified various signatures for future laboratory tests of the inverse-square-law. Subsequent analysis by the E$\ddot{{\rm o}}$t-Wash group~\cite{eotwash} excluded a significant part of the parameter space.

\subsubsection{Thin-Shell Screening}
\label{thinshellcham}

The density-dependent mass immediately results in a further decoupling effect outside sufficiently massive objects, due to the {\it thin-shell} effect. Consider a spherical source of
radius $R$ and density $\rho_{\rm in}$ embedded in a homogeneous medium of density $\rho_{\rm out}$. The corresponding effective minima will be respectively denoted by
$\phi_{\rm min-in}$ and $\phi_{\rm min-out}$. For a sufficiently massive source, the scalar field is oblivious to the exterior matter and is therefore pinned near
$\phi_{\rm min-in}$ in the core of the object. Of course, $\phi$ must deviate substantially from $\phi_{\rm min-in}$ near the surface of the object since $\phi$ must eventually reach the
asymptotic value $\phi_{\rm min-out}$ far away. Thus the gradient in $\phi$ builds up only within a thin-shell of thickness $\Delta R$ below the surface, given by
\be
\frac{\Delta R}{R} = \frac{1}{6\beta M_{\rm Pl}} \frac{\phi_{\rm min-out} - \phi_{\rm min-in}}{\Phi_{\rm N}}\,,
\label{thin}
\ee
where $\Phi_{\rm N}$ is the surface Newtonian potential. In other words, the shell thickness is determined by the difference in $\phi$ values relative to the difference
in gravitational potential between the surface and infinity.  

Since field gradients are essentially confined to the shell, the exterior profile is suppressed by a thin-shell factor:
\be
\phi_{\rm screen} \approx  -\frac{\beta}{4\pi M_{\rm Pl}} \frac{3\Delta R}{R} \frac{Me^{-m_{\rm min-out} r}}{r} + \phi_{\rm min-out}\,.
\ee
The suppression factor of $\Delta R/R$ can alternatively be understood in a more intuitive way as follows. Deep inside the source, the contribution to the exterior profile from
infinitesimal volume elements are Yukawa-suppressed due to the large effective chameleon mass in the core. Only the flux lines from within the
thin-shell can propagate nearly unsuppressed to an exterior probe.

Clearly the thin-shell screening breaks down for sufficiently small objects. Imagine shrinking the source radius keeping the density fixed. Eventually, the cost in gradient energy
to maintain the field difference between $\phi_{\rm min-in}$ and $\phi_{\rm min-out}$ becomes too large, and the scalar field no longer reaches $\phi_{\rm min-in}$ in the core of
the object. In this limit the thin-shell screening goes away, and the exterior profile takes its usual form
\be
\phi_{\rm no\;screen} \approx  -\frac{\beta}{4\pi M_{\rm Pl}} \frac{Me^{-m_{\rm min-out} r}}{r} + \phi_{\rm min-out}\,.
\label{noscreen}
\ee
The criterion for thin-shell screening to be effective is for the right-hand side of~(\ref{thin}) to be $\ll 1$. Clearly, the more massive the source, the easier is this condition satisfied,
as expected. But note that the criterion also depends on the density contrast --- for fixed source, a denser environment implies smaller $\phi_{\rm min-out}$, which
makes the thin-shell condition easier to satisfy. We will see in Sec.~\ref{chamobs} that the thin-shell screening effect leads to striking observational signatures.

\subsubsection{Chameleon Constraints from Tests of Gravity}
\label{chamconstraints}

The tightest constraint on the model comes from fifth-force searches in the laboratory~\cite{fischbach}, which set a limit of $\approx 50\;\mu$m~\cite{adel}
on the fifth-force range for gravitational strength coupling. Modeling the chameleon profile in the E$\ddot{{\rm o}}$t-Wash set-up, taking into account
that torsion-balance measurements are performed in vacuum, it was shown in~\cite{cham1} that, for the inverse power-law potential
$V(\phi) = M^{4+n}/\phi^n$ with $n$ and $\beta$ of order unity, the fifth-force constraint translates into a bound on $M$:
\begin{equation}
M\;\lsim\; 10^{-3}\;\;{\rm eV}\,.
\label{Mlim}
\end{equation}
Remarkably, the upper bound coincides with the energy scale of dark energy: $\Lambda\sim (10^{-3}\;{\rm eV})^4$. This bound not only ensures that the chameleon is consistent with fifth force searches, but it is also sufficient to satisfy all known tests of GR (see Secs.~\ref{newtonian} and~\ref{postnewtonian}), such as searches for WEP violation, Lunar Laser Ranging observations and other tests of post-Newtonian gravity~\cite{cham1,cham2}.

\subsubsection{Observational Signatures}
\label{chamobs}

Because objects of different mass have different effective coupling to the scalar, macroscopic violations of the Equivalence Principle can arise, even in the case where the microphysical lagrangian~(\ref{Scham}) displays no such violations~\cite{cham1,cham2,lamnic}. In particular, test masses that are screened in the laboratory may not be screened in space.
This leads to the striking prediction that satellite tests of gravity, such as the planned MicroSCOPE~\cite{microscope}, Galileo Galilei~\cite{GG} and STEP~\cite{STEP} missions,
might observe violations of the WEP with $\eta \gg 10^{-13}$, in blatant conflict with laboratory constraints. Meanwhile, from~(\ref{noscreen}) the total force (gravitational + chameleon-mediated)
between unscreened particles is a factor of $1 + 2\beta^2$ larger than gravity, which would appear as ${\cal O}(1)$ deviations from $G_{\rm N}$ measured on Earth.

Effective violations of the WEP also results in a host of astrophysical signatures uncovered in~\cite{lamnic}. For example, consider (unscreened) dwarf galaxies in large void regions.
The HI gas in these galaxies is also unscreened, but the stars are screened. This should result in a systematic ${\cal O}(1)$ mismatch in the rotational velocities of these two tracers,
and hence a corresponding mismatch in mass estimates. 

Other interesting observable signals arise if the chameleon couples to the electromagnetic field
\be
\int {\rm d}^4x \sqrt{-g}e^{\beta_{\gamma}\phi/M_{\rm Pl}} F_{\mu\nu}F^{\mu\nu}\,.
\ee
Such a coupling results in photon-chameleon oscillations in the presence of an external magnetic field. The GammeV experiment~\cite{gammev1,gammev2,gammev3}
searches for the afterglow~\cite{afterglow1,afterglow2} that would result from trapped chameleons converting back into photons and has
so far excluded the range $5\times 10^{11} < \beta < 6.4 \times 10^{12}$~\cite{gammev1}. This bound has been further refined recently in a
more complete analysis~\cite{gammev3}. The second-generation experiment, GammeV-CHASE, is currently taking data and will further improve on this bound~\cite{gammev3}. 
Photon-chameleon mixing can also be detected astrophysically, through induced polarization in the spectrum of astrophysical objects~\cite{clare1}
and enhanced scatter in the X/$\gamma$-ray luminosity relation of active galactic nuclei (AGN)~\cite{clare2}. A recent analysis of AGN data shows tantalizing evidence
of this effect~\cite{clare2}.

\subsubsection{Cosmology of Chameleon Models} 

The chameleon scalar field can act as dark energy driving cosmic acceleration. Unfortunately the bound~(\ref{Mlim}) from local tests of gravity
forces the equation of state of the chameleon component to be indistinguishable from $w=-1$, resulting in an expansion history
identical to $\Lambda$CDM. This is closely related to a point made earlier: in order to generate observable deviations from $w=-1$, the chameleon field must be rolling
significantly on a Hubble time, which requires $m_{\rm eff} \sim H_0$ on cosmological scales. But this in turn requires a field displacement of order
$M_{\rm Pl}$ from the local value to the cosmos, which is too costly energetically --- galaxies can no longer maintain their thin shell, leading to gross violations of
local constraints.

Explicitly, the inverse power-law form, $V(\phi) = M^{4+n}/\phi^n$, leads to too small a dark energy contribution for $M\sim 10^{-3}$~eV. In order to account for cosmic acceleration, we must lift this potential by an appropriate constant, which of course has no effect on our earlier analysis of local tests. We can exploit the coincidence of the dark energy scale and the upper bound in~(\ref{Mlim}) to
choose a potential with a single mass parameter~\cite{cham3}
\be
V(\phi) =  M^4 \exp (M^n/\phi^n)\,,
\label{exppot}
\ee
with $M = 10^{-3}$~eV. For $\phi\gg M$, this gives $V(\phi) \approx M^4 + M^{4+n}/\phi^n + \ldots$, which is~(\ref{inversepot}) plus a constant.
Thus, while cosmic acceleration is not explained in chameleon models, it is easily incorporated within its phenomenology.

Let us now turn to the cosmological evolution of the chameleon. In the matter-dominated era,~(\ref{phigen}) reduces to
\be
\ddot{\phi} + 3H\dot{\phi} = -V_{,\phi} -A_{,\phi}\rho_{\rm m}\,,
\label{MD}
\ee
where $\rho_{\rm m}$ satisfies the usual conservation law: $\dot{\rho}_{\rm m} + 3H\rho_{\rm m} = 0$.
In particular, the effective potential for $\phi$ is time-dependent. The behavior of the scalar can be inferred by standard adiabaticity arguments --- as long as $m_{\rm min} \gg H$, the field continuously adjusts to the evolving minimum. This is manifestly satisfied for the potential~(\ref{exppot}) and coupling function~(\ref{coup}),
\be
\frac{m^2_{\rm min}}{H^2} \approx 3\beta\; \Omega_{\rm m}\frac{M_{\rm Pl}}{\phi_{\rm min}} \left\{1+n+n\left(\frac{M}{\phi_{\rm min}}\right)^{n}\right\}\gg 1\,,
\label{mfin}
\ee
since $\phi_{\rm min} \ll M_{\rm Pl}$ for relevant densities, as mentioned earlier. The fact that $m(\phi_{\rm min}) \gg H$ on
cosmological scales today is yet another way of seeing that chameleon dark energy is indistinguishable from a cosmological constant at the background level.

The story is more subtle in the radiation-dominated era, since $\tilde{T}^\mu_{\;\mu} \approx 0$ for a relativistic fluid, and hence
the source term in~(\ref{phigen}) vanishes. This suggests that the field is overdamped and remains frozen until matter-radiation equality, which would lead to inconsistencies with nucleosynthesis if the field started out in the early universe with generic initial displacement $\sim M_{\rm Pl}$. Fortunately, the trace is not always small for a realistic relativistic fluid. As the universe expands and cools, 
matter species in the thermal bath successively become non-relativistic when $m\sim T$. Whenever this happens, $\tilde{T}^\mu_{\;\mu}$ becomes non-zero for about one e-fold of expansion,
until the species in question becomes Boltzmann suppressed, thus driving the field closer to the minimum~\cite{damnord}. The total displacement incurred by ``kicks" from known particles until nucleosynthesis is of order $\beta M_{\rm Pl}$~\cite{cham3}. Thus, for a broad range of initial conditions the field settles to the effective minimum by the onset of nucleosynthesis.

\subsubsection{$f(R)$ Theories}
\label{f(R)}

A class of chameleon theories that has sparked considerable interest over the last few years is the $f(R)$ form~\cite{fR,f(R)2}, 
where the Einstein-Hilbert term in~(\ref{GR_Action}) is replaced by a general function of the Ricci scalar~\cite{fR,f(R)2,f(R)3}:
\be
S = \frac{M_{\rm Pl}^2}{2}\int {\rm d}^4 x \sqrt{-\tilde{g}} f(\tilde{R}) + S_{\rm matter} [\tilde{g}_{\mu\nu}]\,.
\label{Sf(R)}
\ee
See~\cite{f(R)rev,f(R)review} for recent reviews. This action was first introduced by Starobinsky in the context of inflation~\cite{staroinf} and studied further in~\cite{f(R)inf1,f(R)inf2,f(R)inf3}. 
Functions of other curvature invariants have also been considered~\cite{otherinv}, but this generally introduces ghost instabilities~\cite{woodard}.
(Of course general higher-order curvature invariants are harmless from an effective field theory point of view provided they contribute small corrections to Einstein gravity. 
But here one is precisely interested in order-unity deviations from General Relativity.) The only exceptions are functions of the Ricci scalar (as in~(\ref{Sf(R)})) or
functions of the Gauss-Bonnet term~\cite{nojiri}. 

It is well-known that~(\ref{Sf(R)}) is just a scalar-tensor theory in another guise. To see this, introduce a non-dynamical scalar $\psi$:
\be
S = \frac{M_{\rm Pl}^2}{2}\int {\rm d}^4 x \sqrt{-\tilde{g}}\left\{f(\psi) + \frac{{\rm d}f}{{\rm d}\psi}(\tilde{R}-\psi)\right\} + S_{\rm matter} [\tilde{g}_{\mu\nu}]\,.
\label{Sf2}
\ee
Varying with respect to $\psi$ imposes the constraint $\psi = R$, assuming $f''\neq 0$. And since the constraint is algebraic, we can substitute it back into~(\ref{Sf2}),
thereby reproducing~(\ref{Sf(R)}). Furthermore, the field redefinitions $g_{\mu\nu} = ({\rm d}f/{\rm d}\psi) \tilde{g}_{\mu\nu}$ and
$\phi = -\sqrt{3/2} M_{\rm Pl}  \log {\rm d}f/{\rm d}\psi$ reduce~(\ref{Sf2}) to the more familiar form
\be
S = \int {\rm d}^4x\sqrt{-g}\left(\frac{M_{\rm Pl}^2}{2} R - \frac{1}{2}(\partial\phi)^2 - V(\phi)\right) + S_{\rm matter}\left[g_{\mu\nu}e^{\sqrt{2/3}\phi/M_{\rm Pl}}\right] \,,
\label{Scham2}
\ee
where the only remnant of the $f(R)$ choice is encoded in the scalar potential:
\be
V = \frac{M_{\rm Pl}^2 \left(\psi \frac{{\rm d}f}{{\rm d}\psi} - f\right)}{2\left(\frac{{\rm d}f}{{\rm d}\psi} \right)^2}\,.
\ee
The action~(\ref{Scham2}) is precisely of the form~(\ref{Scham}). Theories of $f(R)$ gravity are therefore physically equivalent to a
subclass of chameleon field theories, namely those with $A(\phi)$ given by~(\ref{coup}) for the specific choice $\beta = 1/\sqrt{6}$,
as can be seen from~(\ref{Scham2}). In other words, only for this specific choice of $\beta$ do chameleon theories
allow an $f(R)$ description. 

For completeness, we give the equations of motion in the $f(R)$ language. Varying~(\ref{Sf2}) with respect to the metric
gives the modified Einstein field equations
\be
G_{\mu\nu} + f_{,R} R_{\mu\nu} - \left(\frac{f}{2}-\Box f_{,R}\right)g_{\mu\nu} - \nabla_\mu\nabla_\nu f_{,R} = 8\pi G_{\rm N} T_{\mu\nu}\,,
\label{fREin}
\ee
where $T_{\mu\nu} = -(2/\sqrt{-g})\delta S_{\rm matter}/\delta g^{\mu\nu}$ is the matter stress-energy tensor, and $f_{,R} \equiv {\rm d}f/{\rm d}R$.
Taking the trace of~(\ref{fREin}) yields an equation for the scalar degree of freedom $f_{,R}$:
\be
\Box f_{,R} = \frac{1}{3}\left(R - f_{,R}R + 2f  + 8\pi G_{\rm N}T^\mu_{\;\mu}\right)\,.
\ee

Two popular choices of $f(R)$ are the Hu-Sawicki~\cite{HuSa} and Starobinsky~\cite{Staro} forms, given respectively by
\bea
f(R) &=& R - \frac{a M^2}{1+\left(\frac{R}{M^2}\right)^{-\alpha}}\qquad  {\rm (HS)} \;; \\
f(R) &=& R + aM^2\left[\left(1+ \frac{R^2}{M^4}\right)^{-\alpha/2}-1\right] \qquad {\rm (S)}\,,
\eea
where $a$ and $\alpha$ are positive constants. (For convenience we
have redefined the parameters of~\cite{HuSa} as follows: $M^2 =
m^2/c_2^{1/n}$, $a = c_1c_2^{-1+1/n}$.) 
Both forms satisfy $f(R)\rightarrow 0$ as $R\rightarrow 0$. For the
relevant choices of parameters, however, the entire cosmological
history of our universe lies in the $R\gg M^2$ regime, 
in which both forms have identical limits:
\be
f(R)\approx  R + aM^2 \left[ -1 + \left(\frac{R}{M^2}\right)^{-\alpha}\right] \qquad (R\gg M^2)\,.
\label{largeR}
\ee
In particular, this is the relevant regime to describe the chameleon mechanism.
Using the above dictionary, the corresponding potential for the canonically-normalized scalar field in Einstein frame is 
\be
V(\phi) \approx \frac{a}{2}M^2M_{\rm Pl}^2 \left\{1 - (\alpha+1) \left(\sqrt{\frac{2}{3}} \frac{1}{\alpha a}\frac{\phi}{M_{\rm Pl}}\right)^{\frac{\alpha}{\alpha+1}}\right\}\,.
\label{Vcuspy}
\ee
This potential is finite at $\phi = 0$ (corresponding to $R\rightarrow \infty$ in the $f(R)$ language), but its derivative blows up at this point for any $\alpha > 0$. This is a generic consequence of starting with a nice functional form in the $f(R)$ language --- translating to the scalar-tensor language involves inverting a relation to obtain $R = R(\phi)$, which therefore typically results in a non-analytic
scalar potential.

The pathological form of the potential is indicative that quantum corrections become important near $\phi = 0$~\cite{amolwayneandme}\footnote{We are grateful to Andrew Tolley for helpful discussions on this point.}. 
The scalar potential $V(\phi)$ receives radiative corrections from $\phi$ loops, whose finite part at 1-loop is
\be
\Delta V_{\rm 1-loop} = \frac{V_{,\phi\phi}^2}{64\pi^2} \log\left(\frac{V_{,\phi\phi}}{\mu^{\;2}}\right) \sim \left(\frac{\phi}{M_{\rm Pl}}\right)^{-2 \frac{\alpha+2}{\alpha+1}} \log\left(\frac{\phi^2}{\mu^2}\right) \,,
\label{1loop}
\ee
(As usual there are also divergent pieces that must be fine-tuned to zero.)
This overwhelms the classical part for $\phi \ll M_{\rm Pl}$. Thus scalar configurations in the small-field limit are in the quantum regime in this case. 

The cuspy form of the potential leads to pathologies already at the classical level~\cite{andrei,andreirelated,andreirelated2}: although the static chameleon solutions described earlier only rely on $V_{,\phi}$ becoming large, the fact that $V$ itself remains finite at $\phi = 0$ implies that finite-energy configurations ({\it e.g.} a collapsing star) can dynamically
overshoot the effective minimum and hit the singular point. The singularity at $\phi = 0$ also led~\cite{maedasing} to argue against the existence of
relativistic stars in $f(R)$ gravity, though this conclusion has been questioned recently by~\cite{wayneamol}. What~(\ref{1loop}) shows, however, is that
the classical approximation is not valid for $\phi\ll M_{\rm Pl}$ --- the $f(R)$ dynamics are dominated by quantum effects in this regime.

We should stress that these problems are specific to cuspy potentials of the form~(\ref{Vcuspy}). The inverse power-law form of~(\ref{inversepot}), for instance,
is clearly immune to singularity issues. If a dynamical configuration causes the field to overshoot, the infinite barrier at $\phi = 0$
will stop the field and lead to oscillations around the minimum. (Alternatively, in~\cite{dev,maedasing2} it was shown that adding higher-order curvature
invariants to~(\ref{largeR}) can remove the singular point and push the high-curvature regime to infinity in field space.)
Similarly, the 1-loop contribution~(\ref{1loop}) can be under control over the relevant field range in this case~\cite{amolwayneandme}.

What choice of $f(R)$ leads to inverse power-law potential? In the strong curvature regime, the desired form is
\be
f(R)  = R - aM^2\left[1+ aM^2\left(\frac{R}{M^2}\right)^{\frac{n}{1+n}}\right]  \qquad (R\gg M^2)\,.
\label{inverse}
\ee
This is similar to~(\ref{largeR}), except that the power is positive in this case. Moreover, whereas~(\ref{largeR}) can be an analytic function of $R$, the above 
has branch cuts for any choice of $n$. As mentioned earlier, this is an unavoidable consequence of translating between the $f(R)$ and scalar-tensor forms ---
analytic $V(\phi)$'s map onto non-analytic $f(R)$'s, and vice versa. The stability discussion above therefore implies that non-analytic dependence on the field
variables is forced upon us if we insist on working in the $f(R)$ description. 



\subsection{Symmetron Fields}
\label{symm}

A second mechanism for hiding scalar fields was proposed recently in the context of symmetron field theories~\cite{symmetron,symmetronearlier}. 
As we will shortly, the symmetron technology and building blocks are similar to those of chameleon models. But the physics of the screening mechanism and
its phenomenological consequences are dramatically different. In particular, unlike in chameleon theories, the symmetron has a small mass everywhere
and therefore mediates a long-range force on Earth and in the solar system.

In the symmetron mechanism, the vacuum expectation value (VEV) of the scalar depends on the local mass density, becoming large in regions of low mass density, and small in regions of high mass density. In addition, the coupling of the scalar to matter is proportional to the VEV, so that the scalar couples with gravitational strength in regions
of low density, but is decoupled and screened in regions of high density. 

The starting point is the general chameleon action~(\ref{Scham}). In particular, we assume that the symmetron couples universally to matter fields
through a conformal coupling. The scalar potential is of the symmetry-breaking form,
\be
V(\phi)=-{1\over 2}\mu^2\phi^2+{1\over 4}\lambda\phi^4\,,
\label{Vsymm}
\ee
involving a tachyonic mass scale $\mu$ and dimensionless coupling $\lambda$. Meanwhile, the coupling to matter is governed by
\be
A(\phi) = 1+{1\over 2M^2}\phi^2 + {\cal O}(\phi^4/M^4)\,.
\label{Asymm}
\ee
The field range of interest is restricted to $\phi \ll M$, as we will see shortly, thus higher-order terms in $A(\phi)$ are negligible. The theory described by~(\ref{Vsymm}) and~(\ref{Asymm}) has a $\mathds{Z}_2$ symmetry $\phi\rightarrow -\phi$. The self-interaction potential is the most general renormalizable form consistent with this symmetry. The quadratic coupling to the matter stress tensor is the leading such coupling compatible with the $\mathds{Z}_2$ symmetry. It is non-renormalizable, suppressed by the mass scale $M$. 

The screening mechanism is achieved through an interplay between the symmetry-breaking potential and the universal coupling to matter.
In the case of a non-relativistic source, the effective potential~(\ref{Veffcham}) is given by
\be 
V_{\rm eff}(\phi)={1\over 2}\left({\rho\over M^2}-\mu^2\right)\phi^2+{1\over 4}\lambda\phi^4\,.
\ee
Thus, whether the quadratic term is negative or not, and hence whether the $\mathds{Z}_2$ symmetry is spontaneously broken or not, depends on the local matter density.  
In vacuum, the potential breaks the reflection symmetry spontaneously, and the scalar acquires a vacuum expectation value (VEV) $\phi_0\equiv \mu/\sqrt\lambda$. 
Inside a source of sufficient density, such that  $\rho > M^2\mu^2$, the effective potential no longer breaks the symmetry, and the VEV goes to zero.  

An essential feature is that the lowest order coupling of matter to the symmetron is $\sim\rho \phi^2/M^2$.  Fluctuations $\delta\phi$ around a local background value $\phi_{\rm VEV}$, as would be detected by local experiments, therefore couple as
\be 
\sim{\phi_{\rm VEV}\over M^2}\delta\phi \ \rho\,,
\label{coupling}
\ee
that is, the coupling is proportional to the local VEV.  In high-density, symmetry-restoring environments, the VEV should be near zero, and fluctuations of $\phi$ should not couple to matter. In rarified environments, where $\rho < M^2\mu^2$, the symmetry is broken and the coupling turns back on.  

The case of interest is where  the field becomes tachyonic around the current cosmic density: $H_0^2M_{\rm Pl}^2 \sim \mu^2M^2$. This fixes $\mu$ in terms of $M$,
and hence the mass $m_0$ of small fluctuations around $\phi_0 = \mu/\sqrt{\lambda}$:
\be
m_0= \sqrt{2} \mu \sim \frac{M_{\rm Pl}}{M} H_0\,.
\label{muvalue}
\ee
Local tests of gravity, as we will see, require $M \; \lsim  \;10^{-3}M_{\rm Pl}$. Hence the range $m_0^{-1}$ of the symmetron-mediated force
in vacuum is $\lsim$~Mpc. Meanwhile, if this extra force is to be comparable in strength to the gravitational force in vacuum, then from~(\ref{coupling}) we must impose
$\phi_0/M^2 \sim 1/M_{\rm Pl}$, that is,
\be
\phi_0\equiv \frac{\mu}{\sqrt{\lambda}} \sim \frac{M^2}{M_{\rm Pl}}\,.
\label{vev}
\ee
Together with~(\ref{muvalue}), this implies $\lambda \sim M_{\rm Pl}^4H_0^2/M^6 \ll 1$.
(For $M=10^{-3}M_{\rm Pl}$, in particular, this gives $\lambda = 10^{-102}$ --- the symmetron is extremely weakly coupled.)
We see immediately from~(\ref{vev}) that $\phi_0 \ll M$, hence the field range of interest lies within the regime of the effective field theory,
and ${\cal O}(\phi^4/M^4)$ corrections in~(\ref{Asymm}) can be consistently neglected. 

\subsubsection{Symmetron Thin-Shell Screening}

Symmetron solutions around a source display a thin-shell effect analogous to the chameleon behavior discussed in Sec.~\ref{thinshellcham}. Consider once again the ideal case of a static, spherically-symmetric
source of homogeneous density $\rho >  \mu^2M^2$. For simplicity, we further assume the object lies in vacuum, so that the symmetron tends to its symmetry-breaking VEV far away: $\phi\rightarrow\phi_0$ as $r\rightarrow \infty$. 

For a sufficiently massive source, in a sense that will be made precise shortly, the solution has the following qualitative behavior. Deep in the core of the object, 
the symmetron is weakly coupled to matter, since the matter density forces $\phi\approx 0$ there. Near the surface, meanwhile, the field must grow
away from $\phi = 0$ in order to asymptote to the symmetry-breaking VEV far away. The symmetron is thus weakly coupled to the core of the object,
and its exterior profile is dominated by the surface contribution. In other words, analogously to chameleon models, there is a thin shell screening effect
suppressing the symmetron force on an external test particle. 

The parameter that determines whether a solution is screened or not is
\be
\alpha \equiv \frac{\rho R^2}{M^2} = 6\frac{M_{\rm Pl}^2}{M^2}\Phi_{\rm N} \,.
\label{alp}
\ee
(Recall that $\Phi_{\rm N}$ is the surface gravitational potential.) Objects with $\alpha\gg 1$ display thin-shell screening, and the resulting symmetron-mediated force on a test particle
is suppressed by $1/\alpha$ compared to the gravitational force. Objects with $\alpha \ll 1$, on the other hand, do not have a thin shell --- the symmetron gives an ${\cal O}(1)$ correction to the gravitational attraction in this case. 

\subsubsection{Tests of Gravity and Observational Signatures}

Since the symmetron-mediated force is long-range in all situations of interest, and because the symmetron couples to matter universally,
the relevant tests of gravity are the same that constrain standard BD theories: solar system and binary pulsar observations. A necessary condition
is for the Milky Way galaxy to be screened: $\alpha_{\rm G} \;\gsim\; 1$. Since $\Phi \sim 10^{-6}$ for the galaxy,~(\ref{alp}) implies
\be
M \; \lsim\;  10^{-3}M_{\rm Pl}\,.
\ee
It turns out that this condition is also sufficient to satisfy all current constraints. Indeed, as shown in~\cite{symmetron}, with
$M = 10^{-3}M_{\rm Pl}$ the symmetron predictions for time-delay, light-deflection, perihelion precession of Mercury and Nordvedt effect
are comparable to current sensitivity levels and therefore detectable by next-generation experiments. Note that pushing $M$ to larger
values is also desirable cosmologically, since the range of the symmetron force grows with $M$.
In particular, from~(\ref{muvalue}), $\mu^{-1}\; \lsim \; {\rm Mpc}$ for $M \; \lsim \;10^{-3}M_{\rm Pl}$.

The symmetron is observationally distinguishable from other screening mechanisms. In chameleon theory, as discussed in Sec.~\ref{chamconstraints},
the tightest constraint comes from laboratory tests of the inverse square law. Once this is satisifed, however, the predicted solar system deviations
are unobservably small. In contrast, as mentioned above the symmetron predictions for solar system tests are just below current constraints.
On the other hand, chameleon and symmetron models have in common the prediction of macroscopic violations of the WEP, which
can show up in various astrophysical observations~\cite{lamnic}. In the Vainshtein case, as mentioned in Sec.~\ref{dgpsec}, the DGP model
predicts modifications to the Moon's orbit that are within reach of next generation Lunar Laser Ranging observations, but light-deflection
and time-delay signals are negligible.

\subsection{Theories of Massive/Resonance Gravity}
\label{vain}

A class of infrared modified theories that has spurred a lot of activity involves giving the graviton a small mass or width. This is alluring from a purely theoretical perspective,
because of the formidable challenge devising consistent, ghost-free theories of massive gravity has proven to be. We know how to give a mass to all particles, but doing so consistently for the graviton remains an open problem. Besides the theoretical appeal, the main motivation remains of course the cosmological constant problem. The cosmological constant is the
zero-momentum component of the stress-energy tensor, hence its backreaction depends sensitively on the nature of gravity in the far infrared. 

At the linearized level, all theories of massive/resonance gravity reduce effectively to scalar-tensor theories, even in the limit of vanishing mass, where the scalar is the longitudinal mode of the graviton.
This is the famous van Dam-Veltman-Zakharov (vDVZ)~\cite{vDVZ} discontinuity. The resolution is  the Vainshtein mechanism~\cite{vainshtein} --- non-linearities in the longitudinal dominate in the vicinity of astrophysical sources, and result in its decoupling from matter. Thus GR is approximately recovered in the solar system.

\subsubsection{Fiertz-Pauli Gravity and Its Discontents}
\label{FPgrav}

The perennial challenge for theories of massive/resonance gravity is the avoidance of ghost instabilities. A theory of a single,
Lorentz-invariant massive graviton~\cite{FP} is generically plagued with ghosts. (The story is different for Lorentz-violating massive gravity~\cite{LVmg1,LVmg2,LVmg3};
for simplicity, we focus here on Lorentz-invariant mass terms.) At the linearized level, the Fiertz-Pauli tensor structure~\cite{FP}
is chosen precisely to remove the 6th polarization mode of the graviton, which would otherwise be a ghost. But since this is no symmetry to prevent
this mode from being excited non-linearly, one generically finds a ghost propagating around non-trivial backgrounds. The instabilities of massive gravity were first diagnosed by
Boulware and Deser~\cite{boulwaredeser} in the ADM formalism: integrating out the lapse function and shift vector results in a Hamiltonian that is unbounded from below.
In recent developments, a realization of massive gravity using an auxiliary extra dimension~\cite{aux,claudiaaux} has been argued to have positive definite Hamiltonian. 

The origin of the ghost is most transparent in the Goldstone-St$\ddot{{\rm u}}$ckelberg language~\cite{paolomass,deffrom,degrav}. As in the previous subsection,
the metric fluctuation $\tilde{h}_{\mu\nu} = g_{\mu\nu} - \eta_{\mu\nu}$ is completed to a gauge-invariant object $h_{\mu\nu}$ by including the Goldstone modes.
Focusing on the the longitudinal or helicity-0 mode $\pi$, we have~\cite{ags}
\be
h_{\mu\nu} = \tilde{h}_{\mu\nu} + \eta_{\mu\nu} \frac{\pi}{M_{\rm Pl}} + \frac{2 r_c^2}{M_{\rm Pl}} \partial_\mu\partial_\nu \pi + \frac{r_c^4}{M_{\rm Pl}^2} \partial_\mu \partial^\gamma \pi \;\partial_\nu\partial_\gamma\pi\,,
\label{hel0}
\ee
where the $\eta_{\mu\nu} \pi/M_{\rm Pl}$ term is introduced for convenience to diagonalize the kinetic matrix of $\tilde{h}$ and $\pi$. (This is the linearized version of a conformal transformation to Einstein frame.)
Since $h_{\mu\nu}$ is gauge-invariant, we are allowed to add ${\cal L}_{\rm mass} = \sqrt{-g} g^{\mu\nu}g^{\alpha\beta}(a h_{\mu\alpha}h_{\nu\beta} + b h_{\mu\nu}h_{\alpha\beta})$ to the Einstein-Hilbert lagrangian,
where $a$ and $b$ are constants. Substituting the decomposition~(\ref{hel0}), we risk generating a $(\Box\pi)^2$ term in the action from the square of the
$\partial_\mu\partial_\nu \pi$ term, which would yield a ghost at the quadratic level. But the Fiertz-Pauli structure ($a=-b$) is chosen precisely to remove
this dangerous term. 

We can zoom in on the Goldstone sector by taking the decoupling limit~\cite{ags}: $M_{\rm Pl}\rightarrow \infty$ and $r_c\rightarrow \infty$, keeping
the strong coupling scale $(r_c^{-4}M_{\rm Pl})^{1/5}$ fixed. The resulting action~\cite{ags,paolomass},
\be
S_{\rm FP} = \int {\rm d}^4x \left(3\pi\Box\pi + \frac{r_c^4}{M_{\rm Pl}}\Box\pi \left[(\Box\pi)^2 - (\partial_\mu\partial_\nu\pi)^2\right] + \frac{1}{M_{\rm Pl}}\pi T\right)\,,
\label{SpiFP}
\ee
is clearly well-defined at the quadratic level, as advocated. However, the theory has non-linear instabilities: expanding the interaction term
to quadratic order in perturbations $\varphi$ around a background solution $\Pi(x)$ yields higher-derivative terms of the form $\Box\Pi (\Box\varphi)^2$, which signals a ghost.
One could hope to remove the ghost by adding suitable higher-order terms in $h$ can remove the ghost, but~\cite{paolomass} argued
there is insufficient freedom to do so. However, these conclusions have been questioned recently by~\cite{aux} in the context of auxiliary extra dimensions. 
Another proposal for a non-linear version of massive gravity was proposed recently in~\cite{mukhFP}.

\subsubsection{Dvali-Gabadadze-Porrati Gravity}
\label{dgpsec}

The DGP model~\cite{DGP,DGP2,DGP3} shares many properties of massive
gravity, but avoids some of its pitfalls.
See~\cite{luerev} for a review. In this construction,
our visible universe is confined to a 3-brane embedded in an empty,
4+1-dimensional space-time, where the extra dimension has
infinite-volume. The hallmark of DGP gravity is the inclusion of an
Einstein-Hilbert term on the brane. 
The action, given by
\be
S = \frac{M_5^3}{2} \int_{\rm bulk} {\rm d}^5x\sqrt{-g_5} R_5  + \frac{M_{\rm Pl}^2}{2} \int_{\rm brane} {\rm d}^4 x \sqrt{-g_4}\left (R_4 + {\cal L}_{\rm matter}\right) \,,
\ee
therefore involves two Planck scale, $M_{\rm Pl}$ and $M_5$. The $R_4$ term on the brane gives a large inertia to the graviton and leads to a recovery of a $1/r^2$ force law at short distances.
But since the extra dimension is infinite in extent, GR is not recovered in the infrared --- instead the gravitational force law asymptotes to $1/r^3$ at large distances, corresponding to 5D gravity.
The cross-over scale $r_c$ from $1/r^2$ to $1/r^3$ is set by the bulk and brane Planck scales:
\be
r_c = \frac{M_{\rm Pl}^2}{M_5^3}\,.
\ee
Hence a large hierarchy between $M_5$ and $M_{\rm Pl}$ is necessary to recover the inverse-square law on sufficiently large distances. Setting
$r_c \sim H_0^{-1}$, for instance, requires $M_5 \sim 100$~MeV.

From the point of view of a brane observer, the 5 helicity states of the massless 5D graviton
combine to form a massive spin-2 representation in 4D. In the weak-field limit, the metric fluctuation
on the brane for a given an energy-momentum source is~\cite{DGP,nitti}
\be
h_{\mu\nu}  = \frac{M_{\rm Pl}^{-2}}{-\Box + r_c^{-1} \sqrt{-\Box}} \left( T_{\mu\nu} - \frac{1}{3} \eta_{\mu\nu} T\right)\,.
\label{dgpamp}
\ee
The infrared modification is manifest in the form of the propagator, $1/(-\Box +  r_c^{-1} \sqrt{-\Box})$, which describes a resonance graviton
with tiny width $r_c^{-1}$. (In the parametrization of~(\ref{mparam}), this corresponds to $\alpha=1/2$.) As in massive gravity, the tensor structure of the above
amplitude is different than predicted by General Relativity, even in the limit $r_c\rightarrow \infty$. This is the famous vDVZ
discontinuity~\cite{vDVZ} of massive gravity. As we will review shortly, this discrepancy is an artifact of perturbation theory --- non-linear effects in the
helicity-0 mode near astrophysical sources lead to a recovery of General Relativity. This effect, first conjectured by Vainshtein in the context of
massive gravity~\cite{vainshtein,ddgv,ziour}, has been shown explicitly in DGP~\cite{gruz,por,iglesias,tanaka}.

Since the brane is a codimension-one object, the standard Israel junction conditions lead to a Friedmann equation which is {\it local} on the brane~\cite{cedric}:
\be
H^2 = \frac{\rho}{3M_{\rm Pl}^2} \pm \frac{H}{r_c}\,,
\label{DGPfried}
\ee
where the $\pm$ branches correspond to different brane embeddings in
the Minkowski bulk space-time. The ``+" branch has generated a lot
interest because it leads to cosmic acceleration at late times without
vacuum energy~\cite{cedric,cedric2}. However, it has been established
that this branch of solutions 
suffers from ghost instabilities, first using perturbative arguments~\cite{luty,nicolis,DGPghost} and, more recently, by studying non-linear solutions~\cite{shock,gorbunov,DW,rob}. 
The ``$-$" branch, however, is stable. (To account for cosmic acceleration, the $-$ branch requires a vacuum energy component. Together with the $H/r_c$ correction in~(\ref{DGPfried}), this leads
to an effective dark energy component with $w < -1$~\cite{w<-1}.) This is a generic property in DGP: solutions come in pairs, with one member being continuously
connected to the trivial solution and having stable perturbations, while the other being connected to the self-accelerated cosmological solution and having unstable perturbations.

As in massive gravity, we can zoom in on the helicity-0 sector via a decoupling limit~\cite{luty,ddgv}: $M_{\rm Pl}\rightarrow \infty$ and $r_c\rightarrow\infty$, keeping the strong coupling scale
$(r_c^{-2}M_{\rm Pl})^{1/3}$ fixed. (See~\cite{gigapi} for an argument against the validity of taking a decoupling limit in the DGP case.) In this limit, the theory becomes local on the brane and describes a scalar field $\pi$:
\be
S_{\rm DGP} = \int {\rm d}^4x\left( 3\pi\Box\pi - \frac{r_c^2}{M_{\rm Pl}}(\partial\pi)^2\square\pi +  \frac{1}{M_{\rm Pl}}\pi T\right)\,.
\label{pilag}
\ee
This theory enjoys a Galilean shift symmetry~\cite{luty,nicolis},
\be
\partial_\mu\pi \rightarrow \partial_\mu\pi + c_\mu\,,
\label{gal}
\ee
which is a vestige of the full 5$D$ Lorentz transformations~\cite{IRobs}. Thus, $\pi$ has been dubbed a galileon field~\cite{galileon,deff,nathan,kmouf,galileoncosmo1,galileoncosmo2,galileoncosmo3}.
Even though~(\ref{pilag}) contains higher-derivative interactions, the resulting equation of motion for $\pi$,
\be
\partial^\mu\left(6M_{\rm Pl}\partial_\mu\pi + 2r_c^2 \partial_\mu\pi\square\pi - r_c^2\partial_\mu(\partial\pi)^2\right) = - T \,,
\label{pieom}
\ee
is nevertheless second-order --- all higher-derivative terms cancel out when performing the variation. This is unlike the case of massive gravity, where the variation of~(\ref{SpiFP}) yields a higher-order equation of motion. The $\pi$-action~(\ref{pieom}) was generalized in~\cite{galileon} to include higher-derivative interactions
that preserve the Galilean shift symmetry and yield second-order equations of motion.

The $\pi T$ coupling generates a scalar contribution to the graviton exchange amplitude, which is responsible for the discrepancy between the $1/3$ coefficient
in~(\ref{dgpamp}) and the expected factor of $1/2$ for a massless spin-2 field. As mentioned earlier, this discrepancy persists even as $r_c\rightarrow\infty$. In the presence of a large source, however, the self-interactions terms are important and result in the decoupling of $\pi$ from the source. This Vainshtein screening can be immediately understood from~(\ref{pieom}) by considering the static, spherically-symmetric galileon profile due to a point source. With trivial asymptotic conditions,~(\ref{pieom}) can be integrated to give
\be
\frac{{\rm d} \pi}{{\rm d}r} = \frac{3M_{\rm Pl} r}{4r_c^2}\left(-1+\sqrt{1+\frac{4}{9}\frac{r_\star^3}{r^3}}\right)\,,
\label{dpidr}
\ee
where we have introduced the $r_\star$ scale,
\be
r_\star = (r_c^2r_{\rm Sch})^{1/3}\,,
\label{r*}
\ee
with $r_{\rm Sch}$ denoting the Schwarzschild radius of the source. At short distances, $r\ll r_\star$, the galileon-mediated force is manifestly suppressed compared to the gravity:
\be
\left. \frac{F_\pi}{F_{\rm grav}}\right\vert_{r\ll r_\star} = \frac{|\vec{\nabla}\pi|}{M_{\rm Pl}|\vec{\nabla}\Phi|} =  \left(\frac{r}{r_\star}\right)^{3/2}\ll 1\,.
\ee
Thus, as advocated, the strong interactions of $\pi$ lead to its decoupling near a source, and the theory reduces to Newtonian gravity. The deviation from standard gravity, albeit small in the solar system, is nevertheless constrained by Lunar Laser Ranging observations: $r_c\; \gsim\;  120$~Mpc~\cite{moon,degrav,niayeshghazal}. A comparable bound on $r_c$ has also been obtained by studying the effect on planetary orbits~\cite{battat}. The next generation Apache Point Observatory Lunar Laser-ranging Operation will improve this bound by an order of magnitude~\cite{llr}, thereby probing the interesting regime $r_c\sim H_0^{-1}$. 

At large distances, $r \gg r_\star$, on the other hand, the non-linear terms in $\pi$ are negligible, and the resulting
correction to Newtonian gravity is of order unity:
\be
\left. \frac{F_\pi}{F_{\rm grav}}\right\vert_{r\gg r_\star}  = \frac{1}{3}\,.
\label{weakreg}
\ee
The galileon-mediated force therefore leads to an enhancement of the gravitational attraction by a factor of 4/3. In this far-field regime, the theory reduces to a 
Brans-Dicke theory with $\omega_{\rm BD} = 0$, consistent with~(\ref{dgpamp}). 

Unlike chameleon theories, there are no macroscopic violations of the WEP in this theory~\cite{lamnic}. This is because~(\ref{pieom}) takes the form
of a total divergence $\partial j^\mu_\pi = - T$, which follows from the shift symmetry~(\ref{gal}). This implies a generalized ``Gauss' law"~\cite{nicolis,lamnic}: spherically-symmetric exterior solutions for $\pi$ only depend on the mass enclosed, as can be seen from~(\ref{dpidr}). 

The Vainshtein screening can also be understood by studying perturbations $\varphi$ around a background $\Pi$ profile: $\pi = \Pi + \varphi$.
Expanding~(\ref{pilag}) to quadratic order in $\varphi$ gives
\be
S_{\rm pert} = - \int {\rm d}^4x \left (3\eta_{\mu\nu} + 2 K_{\mu\nu} - 2\eta_{\mu\nu} K \right)\partial^\mu\varphi\partial^\nu\varphi\,,
\label{pipert}
\ee
where $K_{\mu\nu} = -r_c^2\partial_\mu\partial_\nu \Pi/M_{\rm Pl}$ is the extrinsic curvature of the brane in the decoupling limit.
In the non-linear regime, $K \gg 1$, galileon perturbations therefore acquire a large kinetic term and decouple from matter. 

We note in passing that evaluating the above kinetic matrix for the spherically-symmetric background~(\ref{dpidr}) shows that perturbations propagate superluminally
in the radial direction in the range $r_{\rm Sch} \ll r \ll r_\star$~\cite{IRobs}. Superluminal propagation is a generic property of galileon theories~\cite{galileon} and
has recently been shown to hold around exact solutions of DGP~\cite{nicpor}. The existence of superluminality in these models casts doubts on whether they can
be completed in a quantum field theory or perturbative string theory~\cite{IRobs}.

As the above discussion illustrates, much of the DGP phenomenology (and its predictions for tests of gravity) is captured by the decoupling theory. This motivated the authors to~\cite{nathan} to propose a 4$D$ theory of modified gravity, by promoting~(\ref{pilag}) into a fully covariant, non-linear theory of gravity coupled to a galileon field. See also~\cite{deff} for covariantizations of more general galileon theories. While the extension is by no means unique, a canonical choice is
\be
S = \int {\rm d}^4x\sqrt{-g}\left(\frac{M_{\rm Pl}^2}{2}e^{-2\pi/M_{\rm Pl}}R - \frac{r_c^2}{M_{\rm Pl}}(\partial\pi)^2\square\pi + {\cal L}_{\rm matter}[g]\right)\,.
\label{nonlin}
\ee
(The $\pi\Box\pi$ term in~(\ref{pilag}) arises by diagonalizing the kinetic matrix as in~(\ref{hel0}), which also generates the $\pi T$ coupling.) Remarkably, the 4$D$ cosmology that follows from this action reproduces many features of the full-fledged DGP model~\cite{nathan}. The Friedmann equation has two branches of solutions, with stable or unstable perturbations, depending on the sign of $\dot{\pi}$. Moreover, there is a cosmological analogue of the Vainshtein effect: at early times, when the density of the universe is high, non-linear interactions in $\pi$ are important, resulting in the galileon energy density being subdominant compared to the matter or radiation fluid. 

\subsubsection{Degravitation}

One of the underlying motivations for massive/resonance gravity is the cosmological constant problem. This is usually stated as ``Why is the vacuum energy so small?" But since the only observable effect of the cosmological constant is through its backreaction on the expansion of the universe, a more conservative question is ``Why does the vacuum energy gravitate so weakly?" If gravity acts as a high-pass filter, in particular, a large cosmological constant can result in a small expansion rate~\cite{dilute,ADGG,degrav}. The vacuum energy can thus {\it degravitate}. As we will see below, degravitation is closely tied to
massive gravity --- any theory that exhibits degravitation must reduce, at the linearized level, to a theory of massive/resonance gravity~\cite{degrav}. Degravitation is
the gravitational analogue to the well-known screening of charges in a superconducting (or Higgs) phase of electromagnetism~\cite{degrav}. 

A phenomenological modification to Einstein's equations that encapsulates degravitation is
\be
G_{\rm N}^{-1}(\Box r_c^2) G_{\mu\nu} = 8\pi T_{\mu\nu}\,,
\label{filter}
\ee
where Newton's constant, $G_{\rm N}(\Box r_c^2)$, has been promoted to a high-pass filter with cut-off scale $r_c$:
sources with characteristic wavelength $\ll r_c$ gravitate normally, but those with wavelength $\gg r_c$ are degravitated.

Equation~(\ref{filter}) manifestly violates the Bianchi identity ---  an immediate consequence of general covariance ---, and hence cannot be the whole story.
But there is already a problem at the linearized level, $g_{\mu\nu} = \eta_{\mu\nu} + \tilde{h}_{\mu\nu}$, where the Bianchi identity is trivially satisfied.
Rewriting the filter function as $G_{\rm N}^{-1}(\Box r_c^2) \equiv 1 - m^2(\Box r_c^2)/\Box$ and choosing de Donder gauge,
$\partial^\mu \tilde{h}_{\mu\nu} = \partial_\nu \tilde{h}/2$,~(\ref{filter}) reduces to
\be
(\Box - m^2(\Box r_c^2)) \left(\tilde{h}_{\mu\nu} - \frac{1}{2}\eta_{\mu\nu}\tilde{h}\right) =  8\pi T_{\mu\nu}\,.
\label{filterlin}
\ee
Suppose for a moment that $m^2(\Box r_c^2))\equiv m^2$ is constant. Then the problem is obvious: 
the only allowed (Lorentz-invariant) tensor structure of a spin-2 mass term that is ghost-free is the
Fiertz-Pauli form~\cite{FP}: $m^2(h_{\mu\nu} - h\eta_{\mu\nu})$. The mass term in~(\ref{filterlin})
is manisfestly not in this form, hence it cannot describe a consistent theory of a massive spin-2 particle.
Allowing for more general $m^2(\Box r_c^2))$ only exacerbates things, for we can spectrally represent the
graviton propagator,
\be
\frac{1}{\Box - m^2(\Box r_c^2)} = \int_0^\infty {\rm d}M^2 \frac{\rho(M^2r_c^2)}{\Box - M^2}\,,
\label{spec}
\ee
and apply the argument to each massive graviton in the continuum. 

The resolution of this paradox is simple: a massive graviton has 5 polarization states (2 helicity-2, 2 helicity-1 and 1 helicity-0 states),
but~(\ref{filterlin}) is an effective equation that only describes the helicity-2 part of the graviton, obtained by integrating out the other
3 polarization degrees of freedom. For completeness, let us briefly review the proof of~\cite{degrav,giapower}.
The starting point is a generalization of Fiertz-Pauli theory~\cite{FP}, allowing for a momentum-dependent graviton mass:
\be
\left({\cal E}h\right)_{\mu\nu} + \frac{m^2(\Box r_c^2)}{2}(h_{\mu\nu} - \eta_{\mu\nu}h) = 8\pi T_{\mu\nu}\,,
\label{filterlin2}
\ee
where $\left({\cal E}h\right)_{\mu\nu}  = -\Box h_{\mu\nu}/2 + \ldots$ is the linearized Einstein tensor. The diffeormorphism gauge symmetry can
be made explicit by introducing a St$\ddot{{\rm u}}$ckelberg field $A_\mu$~\cite{ags}
\be
h_{\mu\nu} = \hat{h}_{\mu\nu} + \partial_\mu A_\nu + \partial_\nu A_\mu\,,
\label{hat}
\ee
such that  $\delta\hat{h}_{\mu\nu}  = \partial_\mu \xi_\nu + \partial_\nu \xi_\mu$ and $\delta A_\mu = -\xi_\mu$ under gauge transformations. Thus 
$h_{\mu\nu}$ is gauge invariant. This is analogous to electromagnetism in a Higgs (or superconducting) phase in which the photon
becomes a gauge-invariant observable: $A_\mu = \tilde{A}_\mu + \partial_\mu\phi$. 

The idea is to solve for $A_\mu$ and substitute the result back into~(\ref{filterlin2}) to obtain an equation for the helicity-2 modes. First, we substitute the decomposition~(\ref{hat}) into~(\ref{filterlin2}):
\be
\label{pfs}
({\cal E}\hat{h})_{\mu\nu} +  m^2(\Box  r_c^2) \left(\hat{h}_{\mu\nu} - \eta_{\mu\nu} \hat{h} + \partial_{\mu}A_{\nu} + \partial_{\nu} A_{\mu}
- 2 \eta_{\mu\nu} \partial^{\alpha}A_{\alpha}\right)  =  8\pi T_{\mu\nu} \,.
\ee
Taking the divergence of this expression allows us to isolate an equation for $A_{\mu}$:
\be
\label{Aequ}
\partial^{\mu} F_{\mu\nu} =  - \partial^{\mu}  \left(\hat{h}_{\mu\nu} - \eta_{\mu\nu} \hat{h} \right)\,,
\ee
where $F_{\mu\nu} \equiv \partial_{\mu}A_{\nu}  - \partial_{\nu}A_{\nu}$. Note that taking another divergence yields:
$\partial^{\mu} \partial^{\nu}  \hat{h}_{\mu\nu} -  \Box \hat{h} =  0$, hence $\hat{h}$ can be represented in the form
$\hat{h}_{\mu\nu}  = \tilde{h}_{\mu\nu} - \eta_{\mu\nu} \Pi_{\alpha\beta}\tilde{h}^{\alpha\beta}/3$, where
$\Pi_{\alpha\beta} = \eta_{\alpha\beta} - \partial_{\alpha}\partial_{\beta}/ \Box$ is the transverse projector.
Thus $\tilde{h}_{\mu\nu}$ carries two degrees of freedom. Coming back to~(\ref{Aequ}), the solution for $A_{\mu}$ is
\begin{equation}
\label{asolution}
A_{\nu}  =  -  {1 \over \Box} \partial^{\mu}  \left(\hat{h}_{\mu\nu}  - \eta_{\mu\nu} \hat{h} \right) - \partial_{\nu} \Theta\,, 
\end{equation}
where $\Theta$ is an arbitrary gauge function. Substituting~(\ref{asolution}) back into~(\ref{pfs}), and choosing $\Theta$ appropriately, we find
\be
\left(1-\frac{m^2(\Box r_c^2)}{\Box}\right) ({\cal E}\tilde{h})_{\mu\nu} =  8\pi T_{\mu\nu}\,,
\ee
which is the linearized version of~(\ref{filter}). Therefore, as advocated,~(\ref{filterlin}) is an effective equation for the helicity-2 part $\tilde{h}_{\mu\nu}$, after integrating out the extra helicities
of a massive spin-2 representation. 

Hence, any theory that proposes to filter out the cosmological constant must reduce, in the weak-field limit, to a theory of massive or resonance gravity. Of course the reverse is not true ---
not every theory of massive/resonance gravity will exhibit degravitation at the non-linear level. To see degravitation at work at the linearized level,
consider~(\ref{filterlin2}) for vacuum energy, $T_{\mu\nu} = -\Lambda\eta_{\mu\nu}$, and let us focus on the case of constant $m^2$ for simplicity:
\be
\left({\cal E}h\right)_{\mu\nu} + \frac{1}{2r_c^2}(h_{\mu\nu} - \eta_{\mu\nu}h)  = - 8\pi \Lambda \eta_{\mu\nu}\,.
\ee
Note that without the mass term the solution grows unbounded, $h_{ij} \sim \Lambda(t^2\delta_{ij} + x_ix_j)/6$, which is the weak-field version of de Sitter space.
With the mass term turned on, however, the solution is just flat space
\be
h_{\mu\nu} = \frac{\Lambda r_c^2}{3}\eta_{\mu\nu}\,.
\ee
Thus the gravitational backreaction of $\Lambda$ vanishes in a theory of a massive spin-2 particle.

What functional forms for $m^2(\Box r_c^2)$ are allowed? A useful parametrization is the power-law form~\cite{degrav,giapower}
\be
m^2(\Box r_c^2) = r_c^{-2(1-\alpha)}(- \Box)^\alpha\,,
\label{mparam}
\ee
where $\alpha$ is a constant.  First, in order for the modification to be relevant in the infrared, we must impose $\alpha < 1$, otherwise the mass term becomes negligible
as $\Box\rightarrow 0$. Second, in order for the graviton propagator to have a positive definite spectral density, $\rho(M^2) \geq 0$ in~(\ref{spec}), we must require $\alpha \geq 0$.
Otherwise the left-hand side of~(\ref{spec}) vanishes as $\Box\rightarrow 0$, which is inconsistent with $\rho(M^2) \geq 0$. Finally, in~\cite{degrav} it was argued that $\alpha < 1/2$ is necessary in order for the degravitation phenomenon to be effective in a certain decoupling limit of the theory. Hence, the allowed range is
\be
0 \leq \alpha < \frac{1}{2} \,.
\label{alpharange}
\ee
The lower bound corresponds to a hard mass for the graviton, discussed in Sec.~\ref{FPgrav}. The range $\alpha > 0$ corresponds
to a continuum of massive graviton states, which immediately points to extra dimensions of infinite extent. The best known example is the
DGP model (Sec.~\ref{dgpsec}), with $D=5$ bulk space-time dimensions. The DGP mass term, $m^2(\Box r_c^2) = r_c^{-1}\sqrt{-\Box}$,
corresponds to $\alpha = 1/2$, which coincides with the upper bound of~(\ref{alpharange}). Cascading gravity theories (Sec.~\ref{cassec}) generalize the DGP scenario to higher
dimensions. As we will see in Sec.~\ref{cassec}, all such theories correspond to $\alpha \approx 0$ in the far infrared.

\subsubsection{Cascading Gravity}
\label{cassec}

It is natural to ask how the DGP model extends to higher dimensions. Purely from a phenomenological perspective,
it is certainly worthwhile to explore this wider class of infrared modifications to gravity. A second motivation,
however, is the degravitation approach to the cosmological constant problem. The standard DGP model fails to
exhibit degravitation. Although the DGP propagator is of the desired form for degravitating the cosmological constant, unfortunately~(\ref{DGPfried}) does not exhibit degravitation --- 
a large brane tension generates rapid Hubble expansion on the brane. This is consistent with the degravitation condition~(\ref{alpharange}), which requires $\alpha < 1/2$.

On the other hand, theories with $D > 6$ bulk space-time dimensions
all correspond to $\alpha \approx 0$ in the far infrared~\cite{cascade2}. Since gravity is $D$-dimensional in the infrared, the
gravitational potential scales as $1/r^{D-3}$ in this limit. Expanding in terms of massive states,

\be
\Phi(r)= \int_{0}^{\infty} {\rm d}M^2 \rho(M^2) \frac{e^{- Mr}}{r} \sim \frac{1}{r^{D-3}}\,,
\ee
the spectral density must satisfy $\rho(M^2)\sim M^{D-6}$ as $M\rightarrow 0$. Therefore, in the small momentum limit ($\Box\rightarrow 0$),
\be
\lim_{\Box\rightarrow 0} \frac{1}{\Box - m^2(\Box r_c^2)}  \sim  \int_0 {\rm d}M^2 \frac{M^{D-6}}{M^2}\,.
\label{spec2}
\ee
The integral converges for $D > 6$. Therefore, all such theories correspond to $\alpha = 0$ in the infrared. For $D=6$, the integral is
logarithmically divergent, corresponding to $m^2(\Box r_c^2)\sim \log\Box$. 

Realizing these higher-dimensional scenarios has been notoriously difficult. To begin with, the simplest
constructions are plagued by ghost instabilities, even around a flat space background~\cite{sergei,gigashif}.
Secondly, because of the higher-codimension nature of the brane, the 4$D$ propagator is divergent and must be regularized~\cite{geroch,gw,claudia}, usually by
giving the brane a finite thickness. Finally, for a static bulk,  the geometry for codimension $N> 2$ has
a naked singularity at finite distance from the brane, for arbitrarily small tension~\cite{dilute}.
(Interestingly, however, it has been argued that allowing the brane to inflate results in a Hubble rate on the brane
which is {\it inversely} proportional to the brane tension for codimension $N > 2$~\cite{dilute}.)

Recently, it was argued that these pathologies are resolved
by embedding our 3-brane within a succession of higher-dimensional
branes, each with their own induced gravity term, embedded in one another in a flat
$D$-dimensional bulk~\cite{cascade1,cascade2,nonFP}. See~\cite{claudiareview} for a pedagogical review.
In the simplest codimension-2 case, our 3-brane is embedded in a 4-brane within a flat 6$D$ bulk, with action
\bea
\nonumber
S &=& \frac{M_6^4}{2} \int_{\rm bulk} {\rm d}^6x\sqrt{-g_6} R_6 + \frac{M_5^3}{2} \int_{\rm 4-brane} {\rm d}^5x\sqrt{-g_5} R_5 \\
&+& \frac{M_{\rm Pl}^2}{2} \int_{\rm 3-brane} {\rm d}^4 x \sqrt{-g_4}\left (R_4 + {\cal L}_{\rm matter}\right) \,.
\label{Scascade}
\eea
The gravitational force law therefore ``cascades" from 4$D$ ($1/r^2$) to 5$D$ ($1/r^3$) to 6$D$ ($1/r^4$) etc., as we probe larger distances on
the 3-brane, with the cross-over scales set by the ratios  $m_5 = M_5^3/M_{\rm Pl}^2$ and $m_6 = M_6^4/M_5^3$. A similar cascading behavior of the force law was also obtained recently in a different codimension-two framework~\cite{nemanjacascade}. Closely related work on intersecting branes was discussed in~\cite{cascadeothers1}
with somewhat different motivations. See~\cite{nishant,niayeshghazal,nbody1,vpec} for cosmological explorations,
and~\cite{cascadeselfacc} for self-accelerated solutions in this context.

As claimed, the induced (scalar) propagator on the 3-brane is completely regular~\cite{cascade1,cascade2}:
\be
{\cal G}(p)\sim  \frac{1}{p^2 + m_5g(p^2)}\,,
\label{G4exact}
\ee
where 
\be
g(p^2) =\left\{\begin{array}{c}
\frac{\pi}{4}\frac{\sqrt{m_{6}^{2}-p^2}}{\tanh^{-1}\left(\sqrt{\frac{m_{6}-p}{m_{6}+p}}\right)} \;\;\;\;\;\;\;\; {\rm for} \;\; p<m_6\\ \\
\frac{\pi}{4}\frac{\sqrt{p^2-m_{6}}}{\tan^{-1}\left(\sqrt{\frac{p-m_{6}}{p+m_{6}}}\right)}
 \;\;\;\;\;\;\;\;\;\;\; {\rm for}\;\; p> m_{6}\,.
\end{array}\right.
\label{gp}
\ee
If $m_5 > m_6$, then the Fourier transform of~(\ref{G4exact}) for a point source yields a potential that cascades 
from $1/r$ (for $r \ll r_5$) to $1/r^2$ (for $m_5^{-1} \ll r \ll m_6^{-1}$) to $1/r^3$ (for $r\gg m_6^{-1}$). Note the crucial role played by
the 4-brane: in the limit $M_5\rightarrow 0$, one recovers the logarithmic divergence, $ {\cal G}(p) \sim \log p$,
characteristic of codimension-2 branes \cite{geroch}. This can be understood intuitively as follows. Because of the $5D$
Einstein-Hilbert term, the force law must become approximately $5D$ at short distances on the 4-brane.
Hence the lower-dimensional brane behaves effectively as a codimension-1 source, whose
Green's function is therefore regular. These conclusions straightforwardly generalize to arbitrary $D$ dimensions.

The ghost issue is trickier. Perturbing around the flat space solution with empty branes,
one finds that a ghost scalar mode propagates. In the 6D case, this is seen most directly from the tensor
structure of the one-graviton exchange amplitude between conserved sources on the 3-brane
in the UV limit~\cite{cascade1,cascade2}:
\be
{\cal A} \sim T^{\mu\nu} \cdot\frac{1}{\Box}\cdot \left(T'_{\mu\nu} - \frac{1}{3}\eta_{\mu\nu} T'\right) -\frac{1}{6}T\cdot \frac{1}{\Box} T'\,.
\label{cascadeamp}
\ee
We have conveniently separated terms as the sum of a massive spin-2 contribution, with the
well-known 1/3 coefficient, plus a contribution from a conformally-coupled scalar. The
problem is with this scalar mode. The last term in~(\ref{cascadeamp}) is negative, indicating a
ghost mode. This UV behavior is identical to other higher-dimensional scenarios~\cite{sergei,gigashif}.

\begin{figure}[h]
\begin{center}$
\begin{array}{cc}
\includegraphics[width=2.2in]{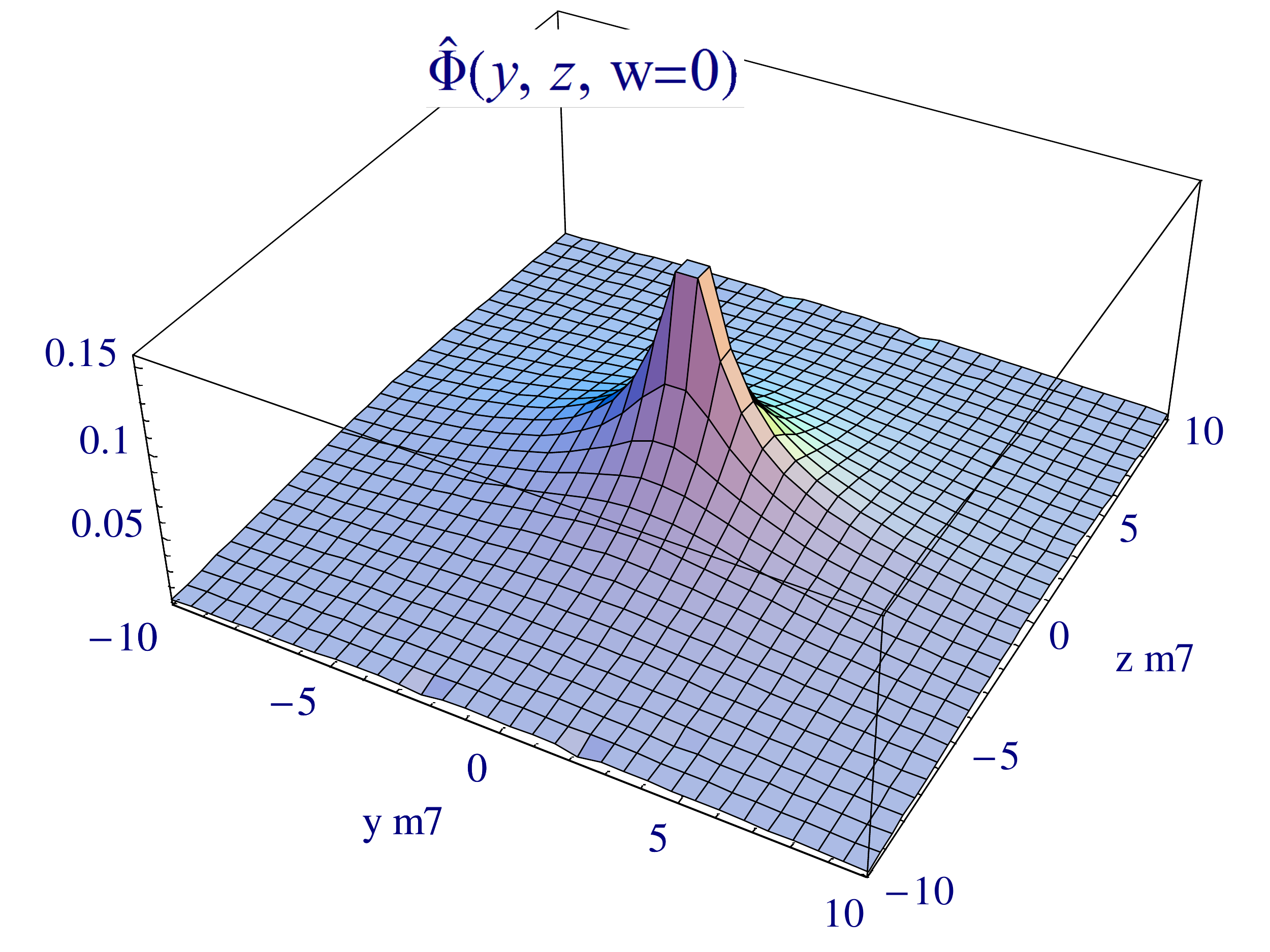} &
\includegraphics[width=2.2in]{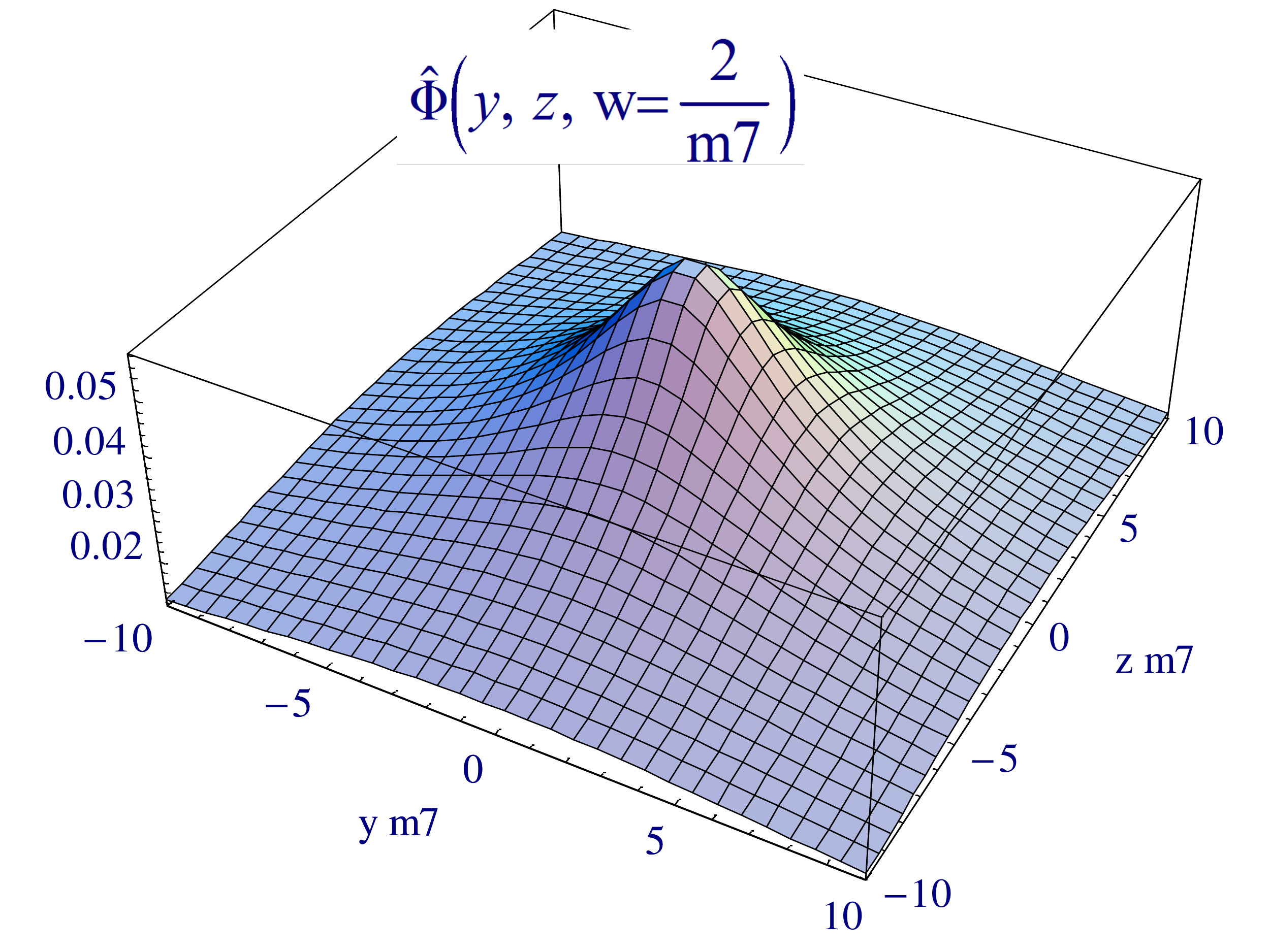}
\end{array}$
\end{center}
\caption{Plot of the $D=7$ cascading solution for the gravitational potential $\hat \Phi(y,z,w)$ resulting from a small tension on the codimension-3 brane. The solution is shown for
$w=0$ and $w=2m_7^{-1}$ in the case where $m_6=m_7$. In terms of the extra-dimensional coordinates $y,z$ and $w$, the codimension-1 brane is located at $w=0$, the codimension-2 brane at $z=w=0$, and the codimension-3 brane at $y=z=w=0$.}
\label{numsol}
\end{figure}

However it was immediately noticed~\cite{cascade1} that the ghost is removed by adding a sufficiently large tension $\Lambda$
on the 3-brane. (Alternatively, the ghost is also cured when considering a higher-dimensional Einstein-Hilbert term localized on the brane~\cite{gigashif,massimo1,massimo2,cascade2}.)
In analogy with a cosmic string in ordinary $4D$ gravity, adding tension to a codimension-2 defect leaves the induced geometry flat but
creates a deficit angle. The exchange amplitude in this case becomes~\cite{cascade1,cascade4}
\be
{\cal A} \sim T^{\mu\nu} \cdot\frac{1}{\Box}\cdot \left(T'_{\mu\nu} - \frac{1}{3}\eta_{\mu\nu} T'\right) +\frac{1}{6\left(\frac{3\Lambda}{2m_6^2M_{\rm Pl}^2} -1\right)}T\cdot \frac{1}{\Box} T'\,.
\label{cascadeamp2}
\ee
The scalar contribution is therefore healthy provided that
\be
\Lambda \geq \frac{2}{3}m_6^2M_{\rm Pl}^2 \,.
\label{lowerbound}
\ee
Note again the essential role played by the 4-brane kinetic term: in the limit $M_5\rightarrow 0$ (keeping $M_{\rm Pl}$ and $M_6$ fixed), we have
$m_6\rightarrow \infty$, and hence~(\ref{lowerbound}) cannot be satisfied. A limitation of the original derivation of~(\ref{lowerbound}) is that it was derived
in a decoupling limit of~(\ref{Scascade}). Recently, this result was confirmed by perturbing the full $6D$ action~(\ref{Scascade}) around a background including tension
on the 3-brane~\cite{cascade4}. The background geometry is flat everywhere, but the extra dimensions show a deficit angle due to the codimension-2 source.
(Just as in DGP, the helicity-0 and conformally-coupled scalar modes contributing to~(\ref{cascadeamp2}) should decouple non-linearly due to the Vainshtein screening effect.)

The $D=6$ framework already exhibits degravitation: as mentioned above, a 3-brane with tension creates a deficit angle in the bulk
while remaining flat. This self-tuning mechanism crucially relies on the extra dimensions having infinite volume: if the dimensions were compact,
the brane tension would have to be tuned against other branes and/or bulk fluxes~\cite{sumrule}. Since the deficit angle must be less than $2\pi$, however,
the tension allowed by the solutions considered in \cite{cascade1,cascade4} is bounded
\be
\Lambda < 2\pi M_6^4\,,
\label{lbound}
\ee
where the $6D$ Planck mass is itself constrained phenomenologically, $M_6 \sim {\rm meV}$. Given the geometrical origin of this bound, it is most likely
an artifact of the codimension-2 case and is expected to be absent in higher-codimension.

Motivated by these considerations,~\cite{cascade3} considered the case of $D=7$ Cascading Gravity, consisting of a 3-brane living on a
4-brane, itself embedded in a 5-brane, together in a 6+1-dimensional bulk. Including tension on the 3-brane, these authors derived a solution for which 
the induced 3-brane geometry is exactly flat. The hope is that in this case brane tension can still degravitate without being limited by a bound analogous to~(\ref{lbound}).
Unlike the case of a pure codimension-3 DGP brane in 6+1 dimensions, where the static bulk geometry has a naked singularity for arbitrarily small tension~\cite{dilute}, here the bulk metric is non-singular everywhere (except, of course, for a delta-function in curvature at the 3-brane location) and asymptotically flat. This smoothness of the solution traces back to the cascading mechanism of regulating the propagator: the presence of parent branes removes the power-law divergence in the $4D$ propagator. 

Figure~\ref{numsol} shows the gravitational potential as a function of the 3 extra-dimensional coordinates, $y,z$ and $w$, with the codimension-1
brane located at $w=0$, the codimension-2 brane at $z=w=0$, and the codimension-3 brane at $y=z=w=0$. Here, $m_7 = M_7^5/M_6^4$ denotes
the cross-over scale from $6D$ to $7D$.  Thanks to the cascading mechanism, which has regularized all potential divergences, the solution shown in Fig.~\ref{numsol}
is finite everywhere. Because the metric depends on 3 spatial coordinates, the analytical results of~\cite{cascade3} was restricted to the weak-field approximation, corresponding
to sufficiently small tension. Ongoing numerical work is aimed at extending these solutions to the non-linear regime of large tension~\cite{claudiaandrewinprogress}.

\section{Tests from Astronomical Observations} 
\label{tests}

The most stringent current tests of gravity  come from laboratory and
solar system tests, spanning millimeter to AU scales.  
In addition, the production of gravity waves predicted by GR is tested
by binary pulsar observations. These tests are summarized above in
Sec.~1 and described in Will's review \cite{Will2001}.  

Astrophysical  probes of gravity on kpc$-$Mpc scales include
galaxy rotation curves, satellite dynamics and other dynamical and lensing properties 
of galaxies and clusters.  On scales of 1 Mpc to over 1 Gpc, tests using the large-scale structure in the universe are being devised using the distribution of galaxies, dark matter (via lensing) and cross-correlations with the CMB. These astrophysical/cosmological tests are the subject of this section. 

Modifications in gravity can also affect the
propagation of gravitational waves. Future gravitational wave
experiments such as LISA can detect gravitational waves from distant
supermassive  black hole pairs in the coalescence phase \cite{GW};
however these tests as well as a variety of other tests of gravity in the strong field regime 
 are beyond the scope of this review. 

\subsection{Early Universe Tests}

The two observational probes of the early universe are Big Bang
Nucleosynthesis (BBN) when the universe was $\sim 10-100$ seconds old
and the Cosmic Microwave Background (CMB) which last scattered when
the universe was $\sim 400,000$ years old.  The abundance of light
elements, in particular Helium and Deuterium, produced during BBN is
very sensitive to the expansion rate $H$ of the universe and the
temperature and time elapsed between the freeze out of neutrons and
protons and BBN (see \cite{Iocco2009} for a review). Its sensitivity
to $H$ and $T$ means that the light element abundance tests the
Friedmann equation, which can be written as: $H^2\sim G g_* T^4$, 
assuming radiation domination and neglecting spatial curvature. Given
our knowledge of $g_*$, the number of relativistic  species at BBN,
this test is usually stated as a constraint on the deviation of the
Gravitational constant $G$ from Newton's value $G_{\rm N}$. Current
constraints on $G$ are somewhat better than 10\%. The value and
constraints on $G$ are quite sensitive to our knowledge of $g_*$ (and
generally assume the current values for non-gravitational constants
such as the fine structure constant).  

The CMB at the epoch of last scattering, $z\simeq 1100$, does not yet
offer as clean a test of the Friedmann equation as BBN. The reason is
that there is a significant degeneracy between $G$ and $H$ unless
precise information on the CMB anisotropy power spectrum is available
\cite{Zahn2003}. A higher value of $G$ leads to a delay in the epoch
of recombination thus causing greater damping of the CMB power
spectrum at high angular wavenumber $\ell$. Using present data on the
CMB temperature  anisotropy power spectrum from the WMAP measurement
plus smaller scale data  \cite{Galli2009} obtain constraints on $G$ at nearly 
the $10\%$ level.  Combining the constraint with BBN, assuming a
constant $G$ between BBN and  $z\simeq 1100$, improves constraints by
a factor of 3. The CMB constraint is limited by the fact that other
parameters, in particular the spectral index of the primordial power
spectrum, can mimic the change in the damping due to deviations in
$G$.   

With the expected measurement of the polarization power spectrum from
the PLANCK satellite, this degeneracy gets broken and the CMB
constraints approach the level of $1$\% \cite{Galli2009}. Thus with
PLANCK and future cosmic variance limited measurements of the CMB
polarization the constraints on $G$ approach the level of current
laboratory measurements.  

\subsection{Large-Scale Structure Tests}

Structure formation in modified gravity in general differs~\cite{Yukawa,Skordis06,Dodelson06,DGPLSS,consistencycheck,Koyama06,fRLSS,Zhang06,Bean06,MMG,Uzan06,Caldwell07,Amendola07,niayeshghazal,nbody1,vpec} 
from that in GR. Perturbative calculations at large scales have shown that it
is promising to connect predictions in these theories with
observations of  large-scale structure (LSS). On small scales 
even larger deviations may occur around galaxy and cluster halos,
depending on  how the theory transitions to GR. 
Nevertheless, carrying out robust tests of MG in practice is
challenging. The two approaches that have been taken are to either
constrain the parameters of a particular model by working out its 
predictions for the growth of structure, or to define 
effective parameters in the spirit of the PPN formalism used
to test GR in the solar system. While both approaches have their 
limitations, we shall see below that there has been recent progress
and upcoming surveys offer great promise for tests of gravity. 

There are three regimes for the growth of perturbations: the
superhorizon regime, the quasi-static Newtonian regime of linear
growth, and  the small scale, nonlinear regime. 
The quasi-static Newtonian regime 
is valid for non-relativistic motions and length scales sufficiently
smaller than the horizon. In this regime (discussed in the next
sub-section) the linearized fluid equations in expanding coordinates
are sufficient to describe perturbations. In the nonlinear regime,
while gravity is still in the weak field limit, density fluctuations
are no longer small and in addition the density/potential fields may
couple to additional scalar fields introduced in MG 
theories. The nonlinear regime is therefore the hardest to describe in
any general way as the nature of the coupling to scalar fields is
theory specific. It may well be the most discriminatory for some
theories as there can be a rich phenomenology ranging from galaxy
cluster to solar system and laboratory scales. 

\subsubsection{Metric and fluid perturbations}

In the Newtonian gauge, scalar perturbations to the metric
are fully specified by two scalar potentials $\Psi$ and $\Phi$:
\begin{equation}
{\rm d}s^2 = -(1+2\Psi)\ {\rm d}t^2 + (1-2\Phi)\ a^2(t)\ {\rm d}{\vec x}^2\,, 
\label{eqn:metric}
\end{equation}
where $a(t)$ is the expansion scale factor. This form for the
perturbed metric is fully general for any metric theory of
gravity, aside from having excluded vector and tensor perturbations
(see \cite{Bertschinger2006} and references therein for justifications). 
Note that $\Psi$ corresponds to the Newtonian potential for
the acceleration of particles, and that in General Relativity
$\Phi=\Psi$ in the absence of anisotropic stresses.  

A metric theory of gravity relates the two potentials above to the
perturbed energy-momentum tensor. We introduce variables to
characterize the density and
velocity perturbations for a fluid, which we will use to
describe matter and dark energy (we will also consider pressure and
anisotropic stress below). 
The density fluctuation $\delta$ is given by
\begin{equation}
\delta({\vec x},t) \equiv \frac{\rho({\vec x},t) -
  {\bar\rho(t)} } {\bar\rho(t)} \,,
\label{eqn:delta}
\end{equation}
where $\rho({\vec x},t)$ is the density and ${\bar\rho(t)}$ is the cosmic
mean density. The second fluid variable is the divergence of the
peculiar velocity 
\begin{equation}
\theta_v \equiv\nabla_j T_0^{\;j}/(\bar{p}+\bar{\rho})={\vec \nabla} \cdot {\vec v}\,, 
\end{equation}
where $\vec v$ is the (proper) peculiar velocity.  
Choosing
$\theta_v$ instead of the vector ${\bf v}$ implies that we have assumed
${\bf v}$ to be irrotational. This approximation is sufficiently
accurate in the linear regime for minimally coupled MG models.

In principle, observations of large-scale structure can directly 
measure the four perturbed variables introduced above: 
the two scalar potentials $\Psi$ and $\Phi$, and the density and velocity
perturbations specified by $\delta$ and $\theta_v$. 
As shown below, these variables are the key to distinguishing 
MG models from GR plus dark energy. Each has a scale and
redshift dependence, so it is worth noting which variables and at what
scale and redshift are probed by different observations. It is
convenient to work with Fourier transforms, such as: 
\begin{equation}
\hat\delta(\vec k,t) = \int d^3 x \ \delta(\vec x,t) \
e^{-i {\vec k} \cdot{\vec x}} \,.
\label{eqn:FT}
\end{equation}
When we refer to length scale $\lambda$, it corresponds to
a statistic such as the power spectrum on wavenumber $k=2\pi/\lambda$. 
We will henceforth work exclusively with the Fourier space quantities 
and drop the $\hat{}$ symbol for convenience. 

The evolution of perturbations can be calculated in the linear 
regime. We follow the formalism and notation of \cite{Ma95}, 
except that we use physical time $t$ instead of conformal time. We are
interested in the evolution of perturbations after decoupling, so we
will neglect radiation and neutrinos as sources of perturbations. 

The superhorizon regime is the most constrained and simplest to
describe. As pointed out by~\cite{Bertschinger2006}, any metric theory
of gravity that also obeys the Einstein equivalence principle must
satisfy a universal evolution equation for metric perturbations. In
the conformal Newtonian gauge, for adiabatic initial conditions, this
evolution is given by:
\begin{equation} 
\ddot{\Phi} - \frac{\ddot{H}}{\dot{H}}\dot{\Phi} + H\dot{\Psi} +
\left(2\dot{H} - \frac{H\ddot{H}}{\dot{H}}\right)\Psi = 0\,. 
\label{eqn:superhorizon}
\end{equation}
The above equation is equivalent to Eqn.~(7) of \cite{HuSa}
which uses a different time variable and the opposite sign 
for $\Psi$. By treating the ratio of metric potential $\Phi/\Psi$ as a constant
parameter of a MG theory, one can solve this equation for a given
background solution $H(t)$. The ISW effect discussed below extends to  
very large scales and probes the superhorizon regime. 

\subsubsection{Quasi-static Newtonian Regime}
In what follows, we will for the most part make the approximation of
non-relativistic motions and restrict ourselves to sub-horizon length
scales. One can also self-consistently neglect time
derivatives of the metric potentials in comparison to spatial
gradients. These approximations will be referred to as the quasi-static,
Newtonian regime. 
Using the linearized fluid equations, the evolution of 
density (or velocity) perturbations can be described by a single second
order differential equation:  
\begin{equation}
\ddot{\delta}+2 H \dot{\delta} + \frac{k^2 \Psi}{a^2}
 = 0 \,. 
\label{eqn:lingrowth}
\end{equation}
With $\delta(\vec k,t)\simeq \delta_{\rm initial}(\vec k) D(k,t)$, 
we can substitute for $\Psi$ in terms of $\delta$ using the 
Poisson equation. Here we write the Poisson equation with the sum of
potentials on the left-hand side, as this is convenient for describing 
lensing and the ISW effect. Using the generalized gravitational
``constant'' $\tilde{G}_{\rm eff}$ we have 
\begin{equation}
\label{eqn:MG3}
k^2(\Psi+\Phi)=-8\pi \tilde{G}_{\rm eff}(k,t)\bar{\rho} \delta \,.
\label{eqn:Poisson}
\end{equation}
Using the two equations above, we obtain for the linear growth 
factor $D(k,t)$: 
\begin{equation}
\ddot{D}+2 H \dot{D} - \frac{8 \pi \tilde{G}_{\rm eff}}{(1+\Phi/\Psi)}
\bar{\rho} \ D = 0 \,. 
\label{eqn:growth}
\end{equation}
From the above equation one sees readily how the combination of ${G}_{\rm eff}$ and $\Phi/\Psi$ alters the linear growth factor. Further, if these parameters have a scale dependence, then even the linear growth factor $D$ becomes scale dependent, a feature not seen in smooth dark energy models. 

\begin{figure}
\includegraphics[width=2.3in]{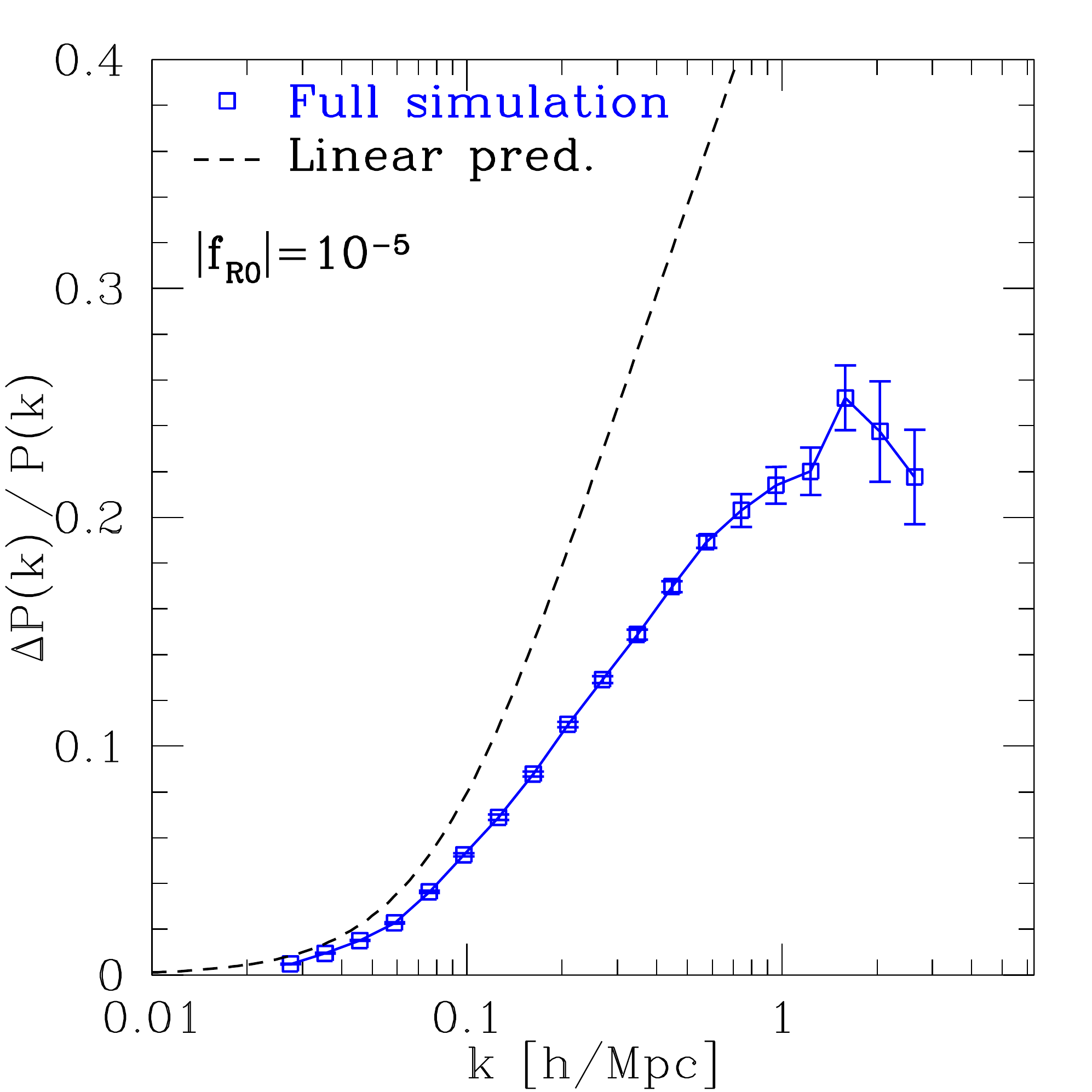}
\includegraphics[width=2.3in]{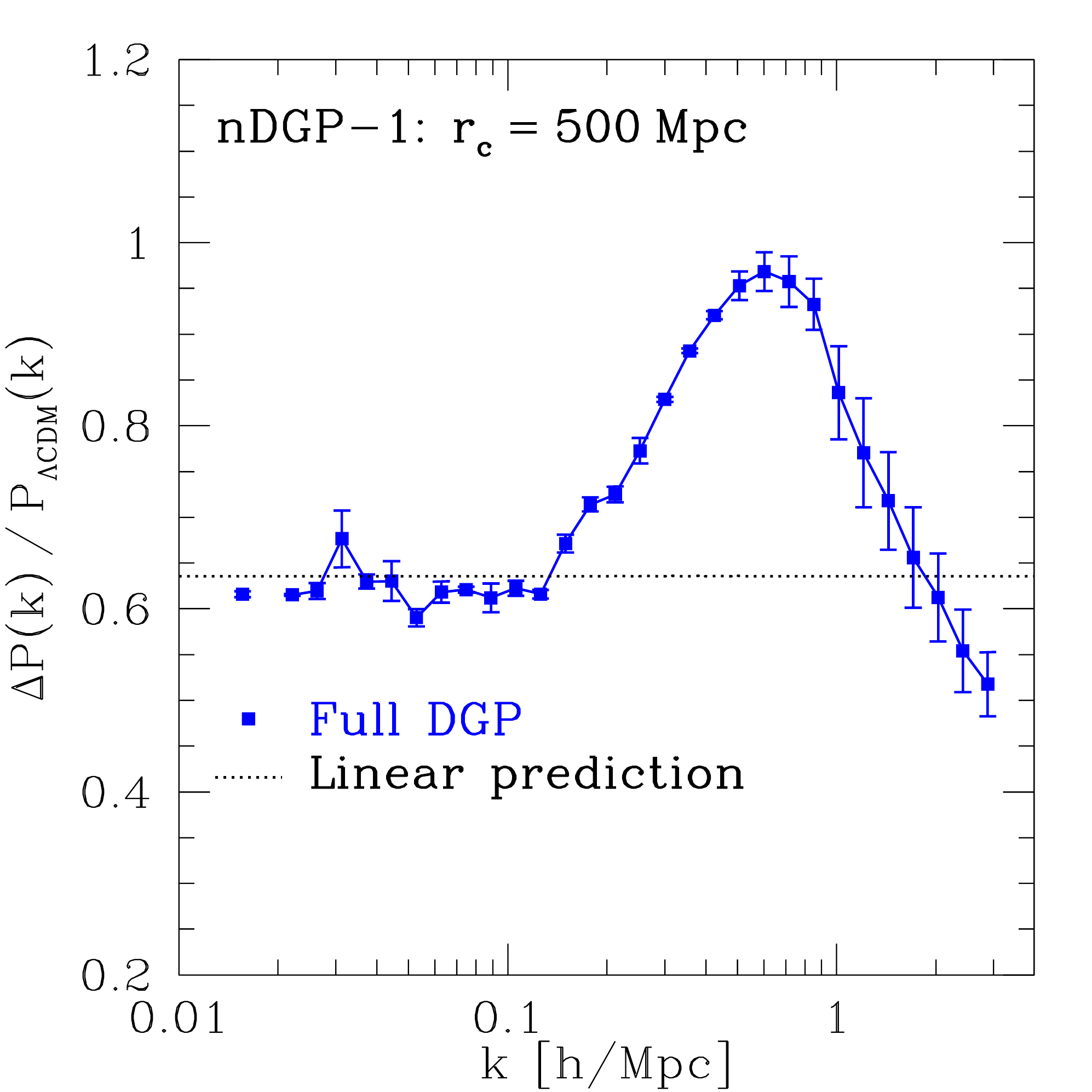}
\caption{Power spectra for $f(R)$ (left panel) and DGP (right panel) theories. 
The fractional deviation from $\Lambda$-CDM are shown for the 
present day linear and nonlinear power spectra \cite{nbody2}.  
At high-$k$ (small scales), nonlinear gravitational clustering and the
screening of massive halos alters the power spectrum. The
limited resolution of the simulations precludes a definitive statement 
about the high-$k$ limit of the power spectra. 
 }
\label{fig:power}
\end{figure}

We will use the power spectra of various observables to describe their scale dependent two point correlations. As an example, 
the 3-dimensional power spectrum of the density contrast $\delta(k,z)$  is defined as  
\begin{equation}
\langle \delta({\vec k}, z) \delta({\vec k'}, z) \rangle = 
(2 \pi)^3 \delta_{\rm D}({\vec k + k'}) P_{\delta\delta}(k,z) \,,
\label{eqn:powerdef}
\end{equation}
where we have switched the time variable to the observable redshift
$z$. The power spectra of perturbations in other quantities are
defined analogously. We will denote the  cross-spectra of two
different variables with appropriate subscripts. For example,
$P_{\delta\Psi}$ denotes the cross-spectrum of the density perturbation $\delta$ 
and the potential $\Psi$.  

Figure \ref{fig:power} shows the linear and nonlinear power spectra 
$P_{\delta\delta}(k,z)$ for $f(R)$ and the normal branch DGP models \cite{nbody2}. 
The dotted curves show the fractional departures of the $f(R)$ linear power spectrum
(left panel) and normal branch DGP (right panel) 
to $\Lambda$CDM. These are simply the ratio of the square of the
linear growth factors. The strong scale dependence is evident in the
$f(R)$ case. In both cases significant deviations 
are evident at wavenumbers $k\simgt 0.1 h$/Mpc.  
The DGP model shown here is the normal branch, which has stronger 
gravitational forces than GR (the self-accelerating branch has weaker
forces than GR). The nonlinear power spectra have smaller deviations
from GR at high-$k$ in part
because the nonlinear contributions are similar for the MG models and
$\Lambda$CDM at sufficiently high wavenumbers. The screening
mechanism within massive halos is the other reason the 
deviations decrease at high-$k$, but the simulations used do not
have the resolution required to fully evaluate this contribution. 

It has been argued that a scale dependence in the MG parameters is
generically expected for scalar-tensor theories~\cite{zhao08}. Provided the interaction can be expressed in the
Yukawa form, one expects a correction to the growth factor that is
quadratic in wavenumber $k$. The authors of~\cite{Amin} have also argued that the
junction condition across the brane in DGP like theories lead to a
linear scale dependent term that can be significant on scales 
approaching the horizon.  

We can also use the relations given above to obtain the linear growth
factors for the velocity and the potentials from $D$. 
The growth factor for the velocity divergence is given 
in the following sub-section, while the Poisson equation
determines the evolution of the potentials.

\subsubsection{Tests using Lensing, ISW and Dynamical Observables}

For tests of gravity, observables that rely directly on the change in
energy or direction of photons are distinct from those that measure
the clustering or dynamics of tracers such as galaxies or galaxy
clusters, which have non-relativistic motions. We summarize the
primary observables that provide tests of gravity on cosmological
scales in this sub-section, and consider galaxy and cluster scale
tests in the following sub-section. We closely follows the treatment 
of Jain \& Zhang  \cite{JainZhang}. 

The CMB power spectrum at angular wavenumber $l$ 
is given by a projection along the line of sight: 
\begin{equation}
C_{TT}(l) = \int dk \int dz \ F_{\rm CMB}(k, l, z)\ j_l[k r(z)] \,,
\end{equation}
where $r$ denotes the comoving angular diameter distance and 
the spherical Bessel function $j_l$ is the geometric term
through which the CMB power spectrum depends on the distance to the last 
scattering surface. The function $F_{\rm CMB}$ 
combines several terms describing the primordial power
spectrum and the growth of the potential up to last scattering. 
We will regard $F_{\rm CMB}$ as 
identical to the GR prediction since we do not invoke MG in the early
universe. 
(See Section 3.1 for a discussion of CMB constraints on $G$ and its
degeneracy with other parameters).  

{\bf The ISW effect: }The CMB anisotropy 
does receive  contributions at redshifts below last scattering, in
particular due to the 
integrated Sachs Wolfe (ISW) effect \cite{ISW}. In the presence of dark
energy or due to modifications in gravity, gravitational potentials
evolve in time and produce a net change in the energy of CMB photons: 
\begin{equation}
\left.\frac{\Delta T}{T}\right|_{\rm ISW}=-\int
\frac{d(\Psi+\Phi)}{dt} \frac{a(z) dz}{H(z)} \,.
\end{equation}
The ISW effect, like gravitational lensing, depends on and probes  the
combination $\Psi+\Phi$. The ISW signal is overwhelmed by the primary CMB at
all scales except for a bump it produces at the largest scales in 
the CMB power
spectrum. For this reason, it is more effectively  measured indirectly, through
cross-correlation with tracers of large scale structure at low redshift. The
resulting cross-correlation signal is a projection of 
$P_{g(\dot\Psi+\dot\Phi)}\left( k,\chi\right)$, the
cross-power spectrum of $(\dot\Psi+\dot\Phi)$ and galaxies (or other
tracers of the LSS such as quasars or clusters).  By cross-correlating
the CMB temperature with galaxy over-density $\delta_g$, the 
ISW effect has been detected at $\la 5\sigma$ confidence level
\cite{ISWmeasurements}  and provides independent evidence
for dark energy, given the prior of a spatially flat universe and GR. It has also provided useful constraints on MG theories as discussed below in Section 3.3. 

\begin{figure}
\includegraphics[width=12cm]{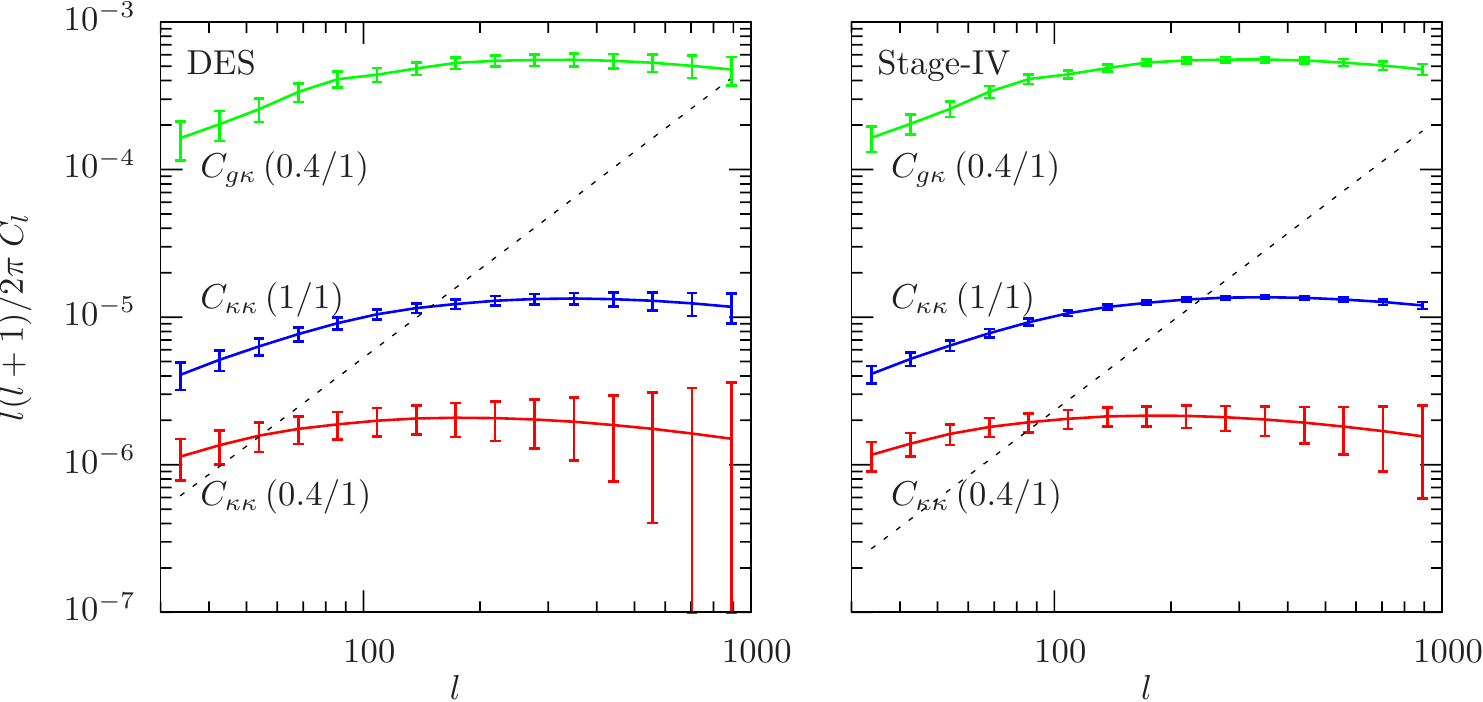}
\caption{
\label{fig:Ckk} 
Examples of the shear-shear and galaxy-shear power spectra for 
the DES (left panel) and a Stage-IV survey similar to 
LSST (right panel) \cite{Guzik09}. 
The upper (green) curves show the galaxy-shear cross power spectrum 
$C_{g\kappa}$, with foreground galaxies at
$z=0.4$ and background galaxies at $z=1$. The lower two curves show  
the shear-shear power 
spectrum  $C_{\kappa \kappa}$ with two choices of redshift bins as
indicated. The error bars include the sample variance and shape noise 
for the two surveys (see Table 1 for survey parameters). 
The shape noise contribution to $C_{\kappa \kappa}$ for $z=1$ is 
shown separately as well (dashed lines). 
}
\end{figure}

{\bf Gravitational Lensing: } Lensing observables are the result of coherent deflections of light by
mass concentrations. For the metric of Eqn. \ref{eqn:metric}, 
the first order perturbation to a photon trajectory is given by
(generalizing for example Eqn.~(7.72) of \cite{Carroll2004}):  
\begin{equation}
\frac{d^2 x^{(1)\mu}}{d\lambda^2} = -q^2 \vec{\nabla}_\perp(\Psi+\Phi)
\,. 
\end{equation}
where $q$ is the norm of the tangent vector of the unperturbed path
and $\vec{\nabla}_\perp$ is the gradient transverse to the unperturbed
path.  This gives the deflection angle formula
\begin{equation}
\alpha_i = -\int \partial_i(\Psi+\Phi) {\rm d}s \,, 
\label{eqn:deflection}
\end{equation}
where $s=q\lambda$ is the
path length and $\alpha_i$ is the $i-$th component of the deflection
angle (a two-component vector on the sky). 
Since all lensing observables are obtained by taking derivatives
of the deflection angle, they necessarily depend only on the
combination $\Psi+\Phi$ (to first order in the potentials). 
The convergence for example is given by
the line-of-sight projection: 
\begin{equation}
   \kappa({\bf \theta}) = \frac{1}{2} \int_0^{z_s} \frac{dz}{H(z)}
    \frac{r(z) r(z_s,z)}{r(z_s)} \nabla_{{\bf 
    \theta}}^2 (\Psi + \Phi)\,,
   \label{eqn:kappa}
\end{equation}
We take the sources to lie at redshift $z_s$. 
The primary observables used in weak lensing are the two-point 
correlations of the observed shapes of galaxies and the
cross-correlation of foreground galaxies with the shapes of background
galaxies. 

The metric potentials are related to the mass distribution  by
(\ref{eqn:Poisson})
so the lensing power spectra can  be expressed in terms of the 
three-dimensional mass power spectrum $P_{\delta \delta}(k,z)$. 
In the small-sky-patch limit  
the Limber approximation \citep{Limber1953} gives 
\begin{eqnarray}
\label{cls1}
C_{\kappa_i \kappa_j}(l) & = & \frac{9}{4} \Omega_m^2 H_0^4   
  \int_0^{\infty} \frac{dz}{H(z)a^2} \, 
\zeta^2(k,z)  
  P_{\delta \delta}(k,z) W_L(z,z_i) W_L(z,z_j),
\end{eqnarray}
where the function $\zeta$ contains the modified gravity parameters: 
$\zeta=G/G_N[(\Phi+\Psi)/\Phi]$ expressed in Fourier space. 
The lensing weight function $  W_L(z,z_k) $
depends on the geometry and the redshift 
distribution of lensed galaxies. 
The three-dimensional wavenumber $k$ is given by $k=l/r(z)$. 
By dividing the galaxy distribution 
into  bins in redshift \citep{Hu1999} a number of auto and cross-spectra can 
be measured. The redshift dependence of these lensing spectra  carries
information about the growth of structure that can test gravity
theories. The galaxy-shear cross-spectrum $C_{g\kappa}$ can be defined
in a similar way to $C_{\kappa\kappa}$: it is proportional to $b\ \Omega_m
\zeta$, where $b$ is the galaxy bias parameter.  $C_{g\kappa}$ is
easier to measure and can be used to test gravity as discussed below. 
Examples of the two lensing spectra are shown in Figure \ref{fig:Ckk} for 
two different survey parameters \cite{Guzik09}. 
Statistical errors are shown for the 
different power spectra -- it is
evident  that if systematic errors can 
be controlled, upcoming surveys will provide percent level
measurements \cite{HoekstraJain}.  

Lensing observables probe the sum of metric potentials -- this follows from the
geodesic equation applied to photons and is therefore true for any metric theory of gravity. Moreover the relation of the sum of metric potentials to the mass  distribution is very close to that
of GR in scalar-tensor theories that we have considered
\cite{Bruneton}. Since the Einstein frame is obtained through a
conformal transformation which cannot alter null geodesics, 
the scalar field does not directly alter the geodesics of light rays. 
Thus the deflection angle formula (\ref{eqn:deflection}) and 
Poisson equation (\ref{eqn:Poisson})  in the form given above are 
essentially unaltered in these scalar-tensory gravity theories. Masses of halos
inferred from lensing are the true masses.  Therefore tests of gravity that
rely solely on lensing measurements  can only be done by using multiple source
redshifts to probe the growth of structure ({\it e.g.},
\cite{Heavens}). It is more useful  to combine lensing with other
observations of large-scale structure.  

{\bf Galaxy and cluster dynamics:} 
MG theories do alter the force experienced by Newtonian test bodies. 
Thus the Newtonian potential $\Psi$ differs from its value in GR in both DGP 
and $f(R)$ theories. It can be stronger by a factor of $4/3$ in a certain regime 
in $f(R)$ gravity (on scales between those of chameleon effects and the Compton
wavelength of the $f_R$ field). This corresponds to the ratio $\Phi/\Psi=1/2$ with the sum 
$\Psi+\Phi$ remaining unaltered as discussed above. Thus for a given 
mass distribution, significant force enhancements can occur. For DGP gravity, similar 
force enhancements occur for the normal branch.  
In particular, the dynamical masses of halos inferred from observations relying on the 
virial theorem or hydrostatic equilibrium  can be significantly larger than the lensing
(or true) masses ({\it e.g.}, \cite{JainZhang,Schmidt2010}).  However in general the ratio of
metric potentials is scale-dependent in MG theories, and force modifications 
can depend on halo mass and environment, so it is not straightforward
to use dynamical and lensing masses to obtain $\Phi/\Psi$. 

Constraints on the Newtonian potential $\Psi$ on small scales are obtained using 
dynamical probes, typically involving galaxy or
cluster velocity measurements. 
On sub-Mpc scales, the Virial theorem for self-gravitating systems in equilibrium 
can be used to constrain $\Psi$ in galaxy 
and cluster halos. Velocity tracers for galaxies include stars and neutral Hydrogen
gas within the halos and satellite galaxies that orbit the outer parts of halos. For galaxy
clusters the tracers are member galaxies and the X-ray emitting hot gas, which is also
mapped using the Sunyaev Zel'dovich (SZ) effect. For relaxed clusters the hot gas is 
assumed to be in hydrostatic equilibrium within the gravitational potential of the halo. 

On large scales, observables depend on the linear growth factor  $D(t)$ given by 
(\ref{eqn:lingrowth}), which determines the 
clustering of matter and is dependent only upon the Newtonian potential $\Psi$. 
The resulting change in the mass power spectrum depends on how much $\Psi$ deviates 
from its GR value and the duration of time this deviation lasts. The results for 
$f(R)$ and DGP models at $z=0$ are shown in
Fig. \ref{fig:power}. However a direct probe of $\Psi$ at a given
redshift is provided by the distortions of galaxy clustering
in redshift space. 

Redshift surveys of galaxies provide statistical measurements of clustering over 
Mpc-Gpc scales. Redshift space distortions in the galaxy power
spectrum arise from motions  along the line-of-sight -- on large
scales these are sensitive to  the linear growth factor for
$\theta_v$, denoted $D_{v}$ here.  
It is related to $D$, the linear density growth factor via the continuity equation as: 
\begin{equation}
D_{v}\propto a\dot{D}=a \beta 
H D \,; \ \  \beta\equiv d\ln D/d\ln a\,. 
\label{eqn:loggrowth}
\end{equation}
Thus redshift space measurements can probes the rate of growth of 
structure. Different combinations of the information from weak lensing and
redshift space galaxy clustering have been used to forecast 
tests of MG models
\citep{Zhang07,acquaviva08,song_dore08,zhao08,zhao09,Guzik09}.  

The line-of-sight component of peculiar velocities cause 
the observable redshift-space power spectrum  $P^{(s)}_{gg}(k, \mu_k)$
to be `squashed' along the line of sight on 
large scales (in the linear regime) and to produce pronounced
`finger-of-God' features on small scales 
(in the nonlinear regime) \citep{Kaiser1987, Hamilton1998}. 
The directional dependence of $P^{(s)}_{gg}$ is given by
$\mu_k \equiv k_{\parallel}/k$, 
which depends on the angle between a wave vector ${\bf k}$ and
the line-of-sight direction. 
Although the picture is more complicated in reality (see \citep{Scoccimarro2004}
for a detailed discussion),  it is a good approximation to decompose the
redshift space power spectrum in terms of three isotropic 
power spectra relating the 
galaxy overdensity  $\delta_g = b\delta$ and peculiar velocities ${\bf v}$:
the galaxy power spectrum $P_{gg}(k)$,
the velocity power spectrum $P_{vv}(k)$ 
and the cross power spectrum $P_{gv}(k)$ as follows 
\citep{Kaiser1987}
\begin{eqnarray}
P^{(s)}_{gg}(k, \mu_k) &=& \left[ P_{gg}(k) + 2 \mu_k^2 P_{gv}(k) + \mu_k^4 P_{vv}(k) \right] F(k^2 \mu^2_k \sigma^2_v)\,, 
        \label{psz:decomp}
\end{eqnarray}
where the term $F(k^2 \mu^2_k \sigma^2_v)$ describes non-linear 
velocity dispersion effects. 
The angular dependence  in the above equation allows us to
obtain the component power spectra from 
$P^{(s)}_{gg}$.  $P_{gg}(k)$, the real space power spectrum of galaxies, 
is the easiest to measure but its interpretation requires knowledge of 
galaxy bias. The pure velocity 
power spectrum $P_{vv}(k)$ has the largest error bars, while 
the cross-spectrum $P_{gv}(k)$ can be estimated more easily. 

The combination of $P_{gv}(k)$, which is proportional to $b D_v$,
with the galaxy-shear cross spectrum $C_{g\kappa}$ is a powerful  
test of gravity \cite{Zhang07}. 
The former probes the growth factor $D_v$ which responds to the
Newtonian potential $\Psi$, while the latter probes the sum of metric
potentials. The ratio is independent of galaxy bias. Hence 
with appropriate redshift binning, these spectra can
constrain the ratio of metric potentials. 
A recent measurement is discussed below. 

\begin{figure}
\includegraphics[width=12cm]{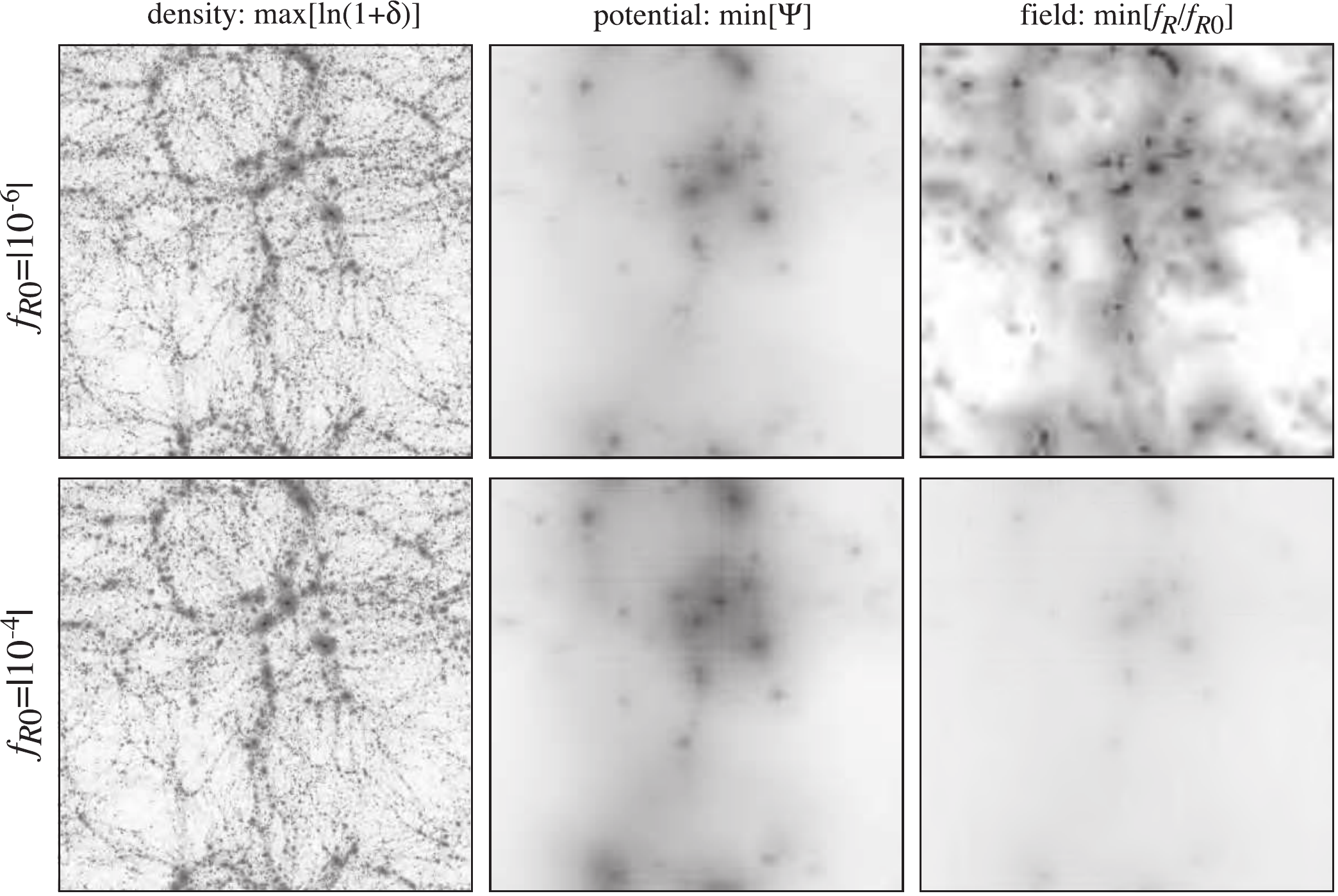}
\caption{Simulation slices showing the density, potential $\Psi$ and $f_R$
  field for two different values of $f_{R0}$ \cite{Oyaizu08}. Along the line of sight 
  through a 2D slice, the maximum of the density and minimum values of $\Psi$ and 
  $f_R$ are shown using a grayscale. In the upper panels the chameleon mechanism 
  is more evident -- it suppresses the deviations (from GR) in the potential gradients in massive 
  structures. In the lower panel the chameleon effect is much weaker due to the choice 
  of a larger value of the $f_{R0}$ parameter. These images show qualitatively the coupling
 between the Newtonian potential and the scalar field in MG gravity theories, and how small 
 scale structure can vary depending on the details of the model.  
}
\label{fig:slice}
\end{figure}

\subsubsection{The Nonlinear Regime: Power Spectra and Halo Properties}

The nonlinear regime can be demarcated in different ways. While it is
standard in cosmology to use the density contrast being close to 
unity as a rule of thumb, for MG we demarcate it using the
breakdown of the linearized equations for the growth of
perturbations. Of particular interest is the coupling of the density
field to a scalar field such as $f_{R}$ for $f(R)$ models. A number
of papers have recently reported simulations that include such a
coupling for both $f(R)$ and DGP gravity \cite{nbody1,Oyaizu08,nbody2,nbody3}. 
Fig. \ref{fig:slice} shows results from
such a study by Oyaizu {\it et al.}~\cite{Oyaizu08}. The density, potential and scalar field $f_R$
are shown for two $f(R)$ simulations. It is evident that the qualitative 
difference in the $f_R$ field leads to a perceptible difference in the potential due
to their coupled evolution. 

\begin{figure}
\includegraphics[width=2.5in]{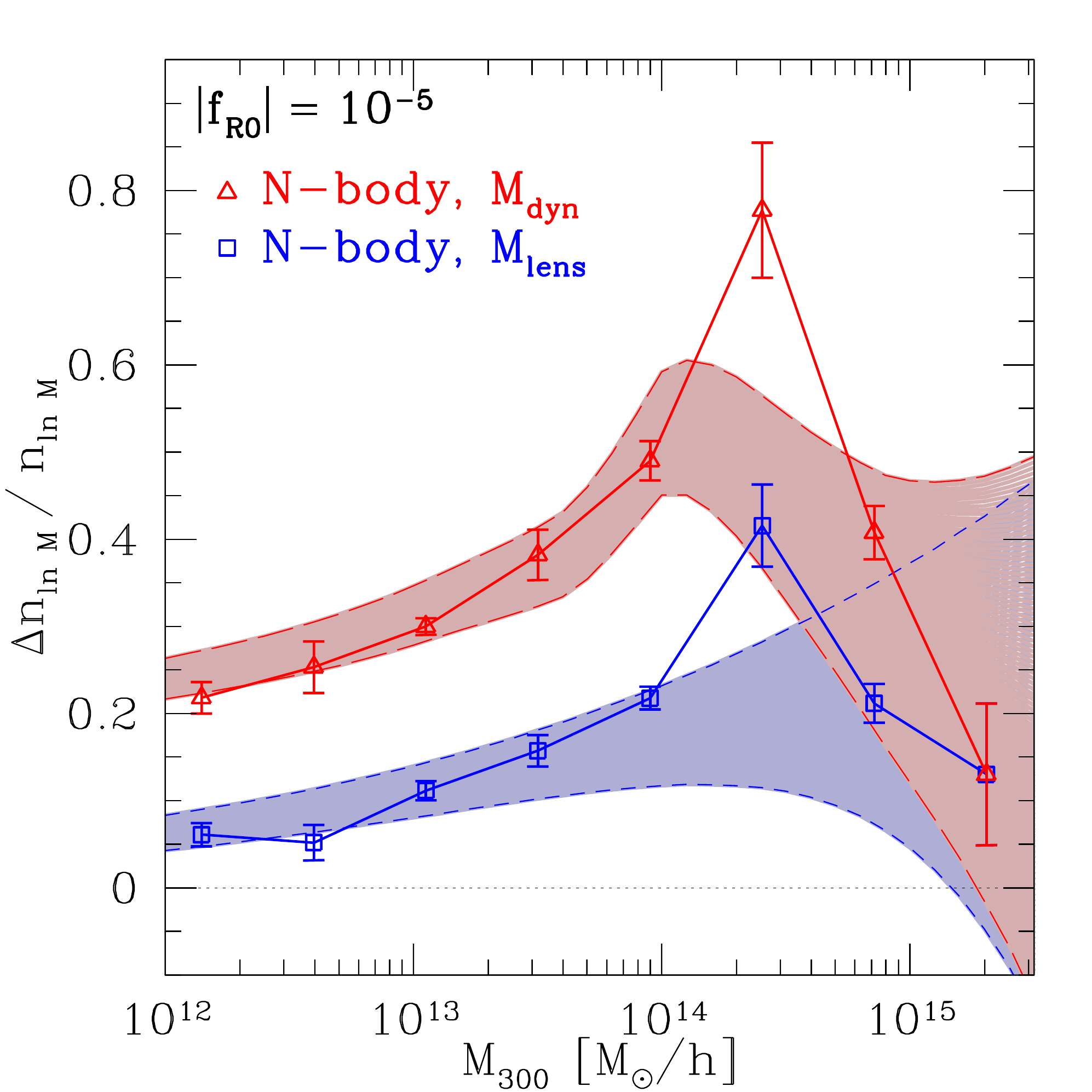}
\includegraphics[width=2.5in]{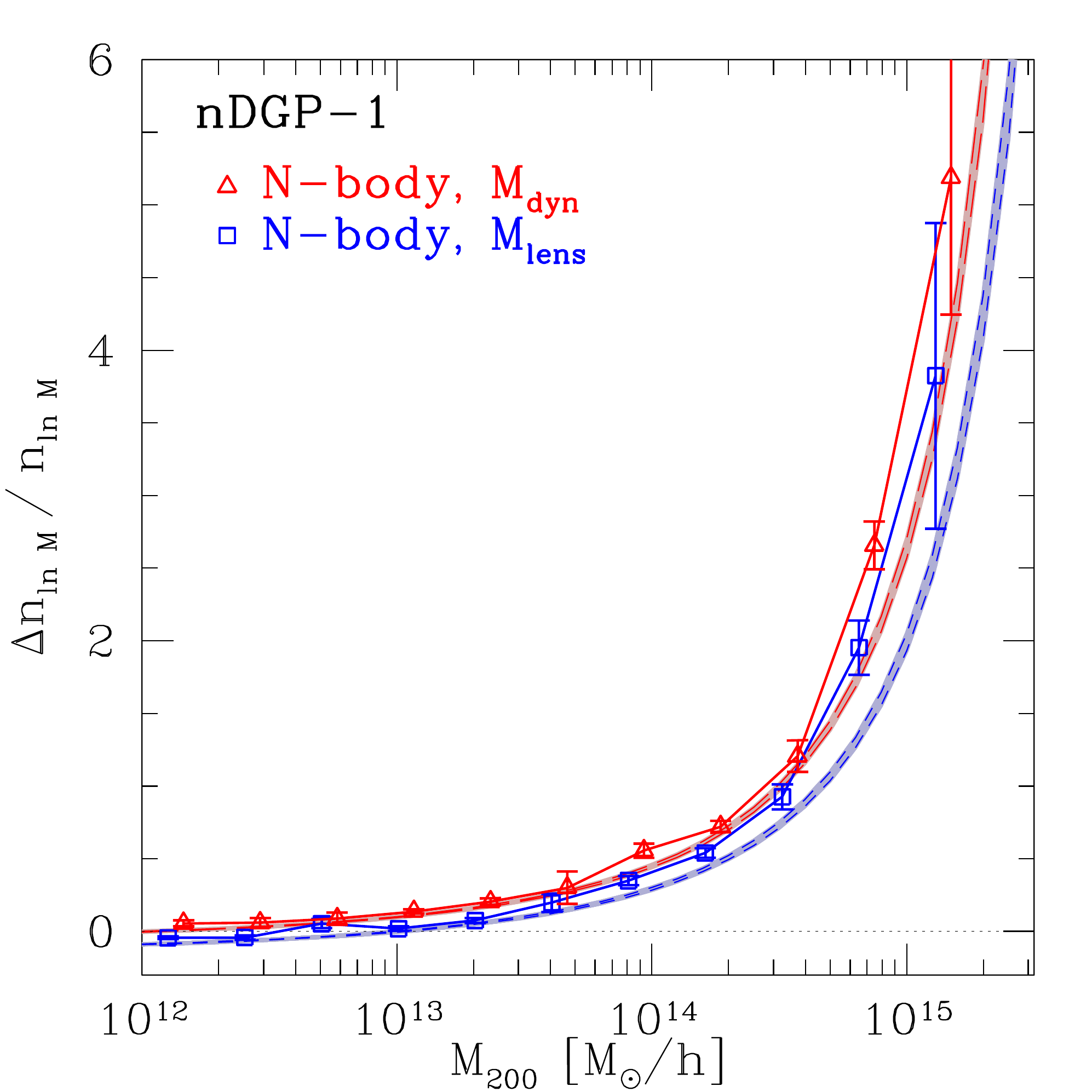}
\caption{Halo mass functions for $f(R)$ (left panel) and normal branch
  DGP (right panel) models. 
The fractional deviation from $\Lambda$-CDM are shown 
using analytical models (shaded regions) and N-body simulations
  (symbols with error bars) \cite{nbody2}. See text for a discussion
  of the two definitions of halo mass. 
 }
\label{fig:massfunction}
\end{figure}

The two primary results from simulations are the nonlinear mass power spectrum 
and the mass function of galaxy and cluster sized halos \cite{nbody2,nbody3,ravi1}. 
For specific models these predictions allow for comparisons to data. Figures~\ref{fig:power} and 
\ref{fig:massfunction} show the measured power spectra and mass functions from 
N-body simulations. Simulations and analytical studies
\cite{Schmidt2010,Smith2009,nbody3} also show the transition from a
modified gravity regime to GR inside massive halos,  
due to the Vainshtein or chameleon mechanisms for DGP and $f(R)$
theories respectively.  Thus we have the possibility
of comparing these MG theories to observations even within and around
galaxy and cluster halos. The advantage of such tests is that order
unity deviations  are expected. 

The mass function of cluster sized halos can show strong deviations due to enhanced 
gravitational forces from the coupling with the scalar field in MG theories. Fig. \ref{fig:massfunction} shows the fractional deviations in the mass function for $f(R)$ and 
DGP models (similar to the power spectra shown in Fig. \ref{fig:power}) \cite{nbody2}. 
For $f(R)$ models significant departures from GR may occur if scales involved in forming clusters are smaller than the Compton wavelength of the scalar $f_R$ but larger than the scale of chameleon effects that screen halos from the modified forces  and drive the theory to GR. 
Theoretical predictions require careful treatment of spherical collapse 
in MG theories as screening effects are dependent on halo mass and
environment \cite{Schmidt2010}. 

For cluster masses approching $10^{15} M_\odot$, the
deviations are significant, from enhancements of tens of percent for $f(R)$ to about a factor of
2 for DGP  models (note that the DGP model shown in Fig. \ref{fig:massfunction} is 
the normal branch DGP, which has enhanced forces, similar to $f(R)$ gravity). 
These departures are driven by the deviation in the linear
growth rate on a scale of $\sim 10$ Mpc, coupled to the exponential dependence of the mass function at the high mass end to this growth rate. In addition, the mass
inferred from  dynamical measurements differs from the true (lensing) mass 
for unscreened halos; this  amplifies the deviations in the mass function 
as shown by the upper set of curves in Fig. \ref{fig:massfunction}. 
The resulting observational
constraints are summarized next.

\subsection{Summary of Current Astrophysical Constraints}

Current tests of gravity on astrophysical scales rely on combining probes of the expansion history with the growth of structure. In the literature so far the observations that have been most effectively used for the distance-redshift relation are: CMB, SNIa and BAOs; and for the growth of structure: the CMB power spectrum, ISW cross-correlations, weak lensing shear correlations and the abundance of galaxy clusters (in the nonlinear regime). 

Perhaps the tightest formal constraints on gravity have been obtained on galactic scales. \cite{Bolton2006} and \cite{Schwab2010} used a combination of strong lensing observations in SDSS galaxies and the dynamics of stars to constrain the ratio of metric potentials. They find the ratio to be consistent with unity, to better than 10\%. 

Cosmological parameter analyses have been performed by several authors
to test gravity in the linear regime of perturbations
\cite{Song2007,niayeshghazal,Schmidt2009,Rapetti2009,Giannantonio,Lombriser,Bean2010,Daniel2010,Zhao2010}. 
In the linear regime (scales $\gsim 10$ Mpc) current  
tests show consistency with GR. For the DGP scenario, \cite{Lombriser}
have  obtained  stringent constraints on the self-accelerating branch
DGP model. They also find   that both normal and self-accelerating
branches of DGP require an explicit cosmological  constant to be viable. 
However, generalized braneworld models such as cascading gravity evade these
observational constraints. 

On scales above 10 Mpc a very recent test of GR  
\cite{Reyes2010} comes from comparing galaxy-velocity and galaxy-shear
cross-correlations (discussed above in Section 3.2.3) from the
SDSS. The $E_G \sim C_{g\kappa}/P_{gv}$ 
parameter introduced by \cite{Zhang07} is a ratio of cross-correlations that is essentially independent of galaxy bias for suitable choices of galaxy redshift distributions. 
It is estimated to be $\sim 0.4$, 
consistent with its value in GR: $E_G = \Omega_m(z=0)/\beta(z)$, where
$\beta$ is the logarithmic rate of growth parameter introduced above
in Eqn. \ref{eqn:loggrowth}. 
The $20$\% level measurement of $E_G$ by \cite{Reyes2010} spans 
scales of 10-50 Mpc at redshift $z\simeq 0.3$. 

The abundance of galaxy clusters from X-ray observations has been used
to constrain the growth factor \cite{Rapetti2009} and specific $f(R)$ models \cite{Schmidt2009}. The mass function can be significantly enhanced at large masses for $f(R)$ and DGP models 
(see discussion in the previous sub-section). Using
information on the mass function requires nonlinear regime model
predictions that include chameleon dynamics, hence constraints are
specific to particular models. \cite{Schmidt2009} constrain the
Compton wavelength to be smaller than $\sim 50\ $Mpc, or equivalently, 
for the present day field amplitude to be $f_{R0}<2\times 10^{-4}$ for
their $f(R)$ model.  A more recent analysis by \cite{Lombriser2010} 
combines large-scale structure information with galaxy cluster
abundances to find comparable constraints.  

Thus current tests of gravity find no indications of departures from
GR. The tests in the literature so far are not very stringent in the linear regime, 
where model-independent constraints on MG parameters are feasible. Tests
on smaller scales have been used to test GR by measuring the ratio of dynamical 
to lensing masses. These 
 tests have constrained the ratio of metric potentials at the 
tens of percent level. Parameters of specific models of $f(R)$ and DGP
gravity have also been constrained. These tests are restricted
to narrow ranges in mass/length scale and redshift, and are far less 
stringent than laboratory or solar system measurements. 
These limited tests are
nevertheless a sign of progress in both theory and observational
analysis -- a decade ago virtually no tests of GR were available on
astrophysical  scales.   
We discuss next the prospects for tests of gravity from upcoming surveys. 

\subsection{Upcoming Surveys}

\begin{table}
\begin{tabular}[b]{r  c c c c c c }
\hline
	& $f_{\mathrm{sky}}$ & $n^{\mathrm{2d}}_{\mathrm{g}}$ & $\bar{n}_{\mathrm{g}}$ & $\left< z \right>$ & $z_{\mathrm{max}}$\\
\hline 
DES        &  $ 5000 $  	& $15$     & -      & $0.7$ & - \\ 
Stage-IV & $ 20000 $	& $30$      &-   	 & $1.2$ & - \\ 	
BOSS      &  $ 10000 $  	&  $0.05$   & $1.1 \times 10^{-4}$   & - & $0.7$   \\ 
BOSS-II   &  $ 20000 $	 & $0.14$    & $1.1 \times 10^{-4}$   & - & $1.1$    \\ 	
\hline
\end{tabular}
\caption{
\label{surv2}
Parameters of imaging and spectroscopic surveys: 
sky coverage $f_{\mathrm{sky}}$ in sq. degs., galaxy surface 
density $n^{\mathrm{2d}}_{\mathrm{g}}$ per sq. arcmin., the three-dimensional 
galaxy number density $\bar{n}_{\mathrm{g}}$ per Mpc$^3$, the mean redshift $\left< z \right>$, 
and the maximum redshift $ z_{\mathrm{max}}$. 
}
\end{table}

Planned surveys that will advance low-redshift cosmological measurements 
include multi-color imaging surveys and spectroscopic surveys of selected galaxies. 
We consider two planned imaging surveys: 
the Dark Energy Survey (DES) \citep{des} which is expected to begin data 
acquisition in 2011, and a generic Stage-IV survey 
\citep{detf,lsst,jdem,euclid} whose 
example is the LSST survey \citep{lsst}. A more complete list of
two generations of upcoming imaging surveys, 
and the redshift range covered, is: 
\begin{itemize}
\item DES, PS1, HSC (2010$-$). $0<z\simlt 1$.
\item LSST, JDEM, EUCLID (2016$-$). $0<z\simlt 3$. 
\end{itemize}

\begin{figure}
\includegraphics[width=12cm]{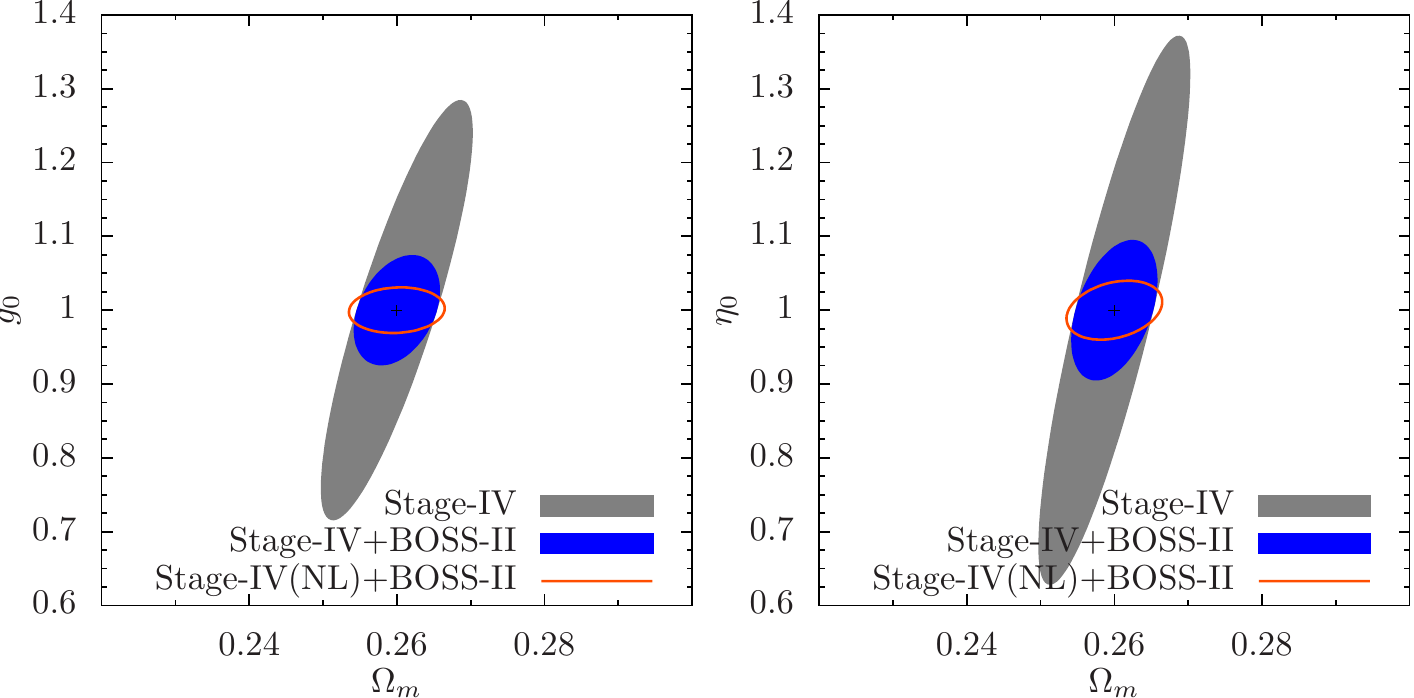}
\caption{Fischer constraints on modified gravity parameters from
  planned Stage IV surveys: LSST and BOSS-II  \cite{Guzik09}. 
  The parameters on the y-axis are $g_0=G/G_N$ and 
  $\eta_0=\Phi/\Psi$. The gray contours show the 68\% confidence region using
  just imaging data while the blue contours combine imaging and spectroscopic 
  (lensing and dynamical) information. The inner red contours also use information 
  in the nonlinear, small-scale regime. 
}
\label{fig:Forecasts}
\end{figure}

\begin{figure}
\includegraphics[width=12cm]{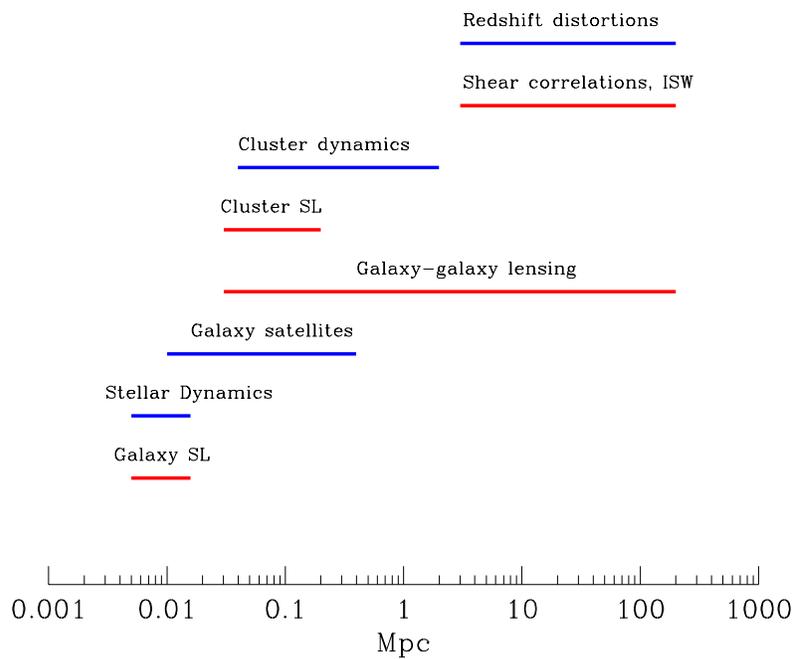}
\caption{Tests of gravity at different length scales. Red lines shows observations
that probe the sum of metric potentials via weak and strong gravitational lensing or
the ISW effect. Blue lines show dynamical measurements that rely on the motions
of stars or galaxies or other non-relativistic tracers. With upcoming surveys, scales 
of 100s of Mpc will become accessible, while the accuracy of measurements will 
improve on all scales. 
}
\label{fig:GravityScales}
\end{figure}

The surveys are characterized by 
sky coverage $f_{\mathrm{sky}}$, 
surface number density of lensed galaxies $n^{\mathrm{2d}}_{\mathrm{g}}$ and 
the galaxy redshift distribution. The sky coverage for
the DES is taken to be 
$5000$ sq. degs. and for the Stage-IV survey $20000$ sq. degs.  
The number density and mean redshifts of the surveys are given  
in Table \ref{surv2}; see \cite{JainZhang} for the detailed redshift 
distribution used and other parameters.  

In order to measure the redshift-space power spectrum $P^{(s)}_{gg}$  
we consider spectroscopic surveys of galaxies. 
While primarily designed to measure
the distance-redshift relation and $H(z)$ using the baryon acoustic
oscillations in galaxy power spectra, they will provide improved 
measurements of $P_{gg}$, $P_{gv}$, $P_{vv}$ on large scales. 
Some upcoming spectroscopic galaxy surveys are: 
\begin{itemize}
\item LAMOST, 
WiggleZ, HETDEX, BOSS (2010$-$). $0<z\simlt 3$. 
\item JDEM, EUCLID, BigBOSS (2016$-$). $0\simlt z\simlt 3$. 
\end{itemize}

As an example of an upcoming survey, we consider the Baryon Oscillation
Spectroscopic Survey (BOSS) \citep{boss} which will target Luminous Red 
Galaxies (LRG) up to redshift $z \sim 0.7$ and will cover a quarter of the sky.
It will obtain spectra of $1.5\times10^6$ galaxies. 
In addition to the BOSS survey  we consider a futuristic version 
(dubbed BOSS-II here; see also the proposed BigBOSS survey \cite{BigBOSS}) 
with double the sky coverage compared to 
BOSS (to keep up with the sky coverage of Stage-IV survey), 
the same galaxy number density, and extending to 
redshift $z=1.1$. Fig. \ref{fig:Ckk} shows the expected accuracy in 
lensing spectra from the DES and Stage IV surveys. 
Fig. \ref{fig:Forecasts} shows the level of tests of gravity achievable 
with planned Stage IV surveys \cite{Guzik09} using lensing and redshift space 
power spectra. The constraints on the Gravitational constant and ratio of 
metric potentials are shown using different combinations of survey data. 

There is another kind of redshift survey based on 21 cm radiation 
from neutral Hydrogen at high redshift. The most ambitious planned
survey is the square kilometer array (SKA) \cite{ska}(2015$-$). 
SKA has the potential to 
detect $\sim 1$ billion galaxies over $0<z\simlt 1.5$, with a deeper
survey extending to $z\sim 5$, through 21cm line
emission of neutral hydrogen in galaxies. If successful, it will 
provide high precision measurements of the distance-redshift relation
through BAO's \cite{Abdalla05}, and tests of MG \cite{Zhang07} through: 
weak lensing, velocity measurements 
through redshift distortions of galaxy clustering, 
and ISW measurements through CMB-galaxy cross-correlations.

X-ray and SZ cluster surveys (e.g. ACT \cite{ACT} and SPT \cite{SPT}) 
aim to measure the mass function
of galaxy clusters out to $z\sim 1$. Cosmological
applications will depend on supplementary optical data to get
photometric redshifts of the detected clusters. The comparison with
lensing masses is essential for the mass calibration, but it will
also provide a test of gravity by comparing dynamical and lensing
masses. 

CMB temperature and polarization maps provide high-z
  constraints and also measurements of the ISW effect and CMB lensing,
  which are probes of $\Psi+\Phi$ at lower redshift. 
In addition to the all-sky PLANCK mission from space, a number of
  surveys from the ground will provide polarization maps at high
  angular resolution. 

We have indicated the approximate redshift range over
which these surveys will provide accurate measurements. 
Different observables that overlap in
redshift and length scale in the range 
$z\simeq 0.3-1$ and at scale $\lambda\simeq$ 10 to several 100
Mpc can be compared in the linear and quasilinear regimes of
structure formation. It is likely that MG effects are significant on
these scales. As emphasized above,  on smaller scales of
order a Mpc or less, even larger deviations may be expected. 
Fig. \ref{fig:GravityScales} shows the length scales expected to be 
probed at low redshift by lensing and dynamical probes. 

We can expect that theoretical developments
will continue to influence observational  
approaches since no complete formalism exists to provide model
independent tests of MG theories on astrophysical scales. It is only
for linear fluctuations in the quasi-static, Newtonian regime that 
the effective Gravitational constant and the ratio of metri potentials
are useful effective parameters. New tests
in the nonlinear regime will require detailed models and simulations
that incorporate the screening mechanisms for specific
theories. Combinations of laboratory and solar system tests with
astrophysical tests are also promising probes of screening mechanisms. 
Finally, we note that we have not explored the feasibility of distinguishing 
modifications of gravity from general dark energy scenarios. Indeed, it is likely that
dark energy with small scale clustering that allows for 
any desired anisotropic stress and pressure will
not in practice be distinguishable from the kinds of MG theories we have discussed (e.g. 
\cite{kunz,JainZhang}). 

\section{Discussion}
\label{discuss}

We have reviewed modified gravity theories with interesting cosmological 
consequences. We focused 
on higher dimensional approaches and scalar-tensor theories, such as chameleon/$f(R)$ and symmetron theories. We classified
these models in terms of the screening mechanisms that enable such theories 
to approach general relativity on small scales. 
We described general features of the modified Friedman equation in such
theories and how they may be distinguished experimentally. The second
half of this review discussed experimental test of gravity in light of
the new theoretical approaches. We introduced a variety of
astrophysical tests that are likely to be valuable in testing gravity 
theories. Recent progress in modeling the growth of perturbations in 
MG theories and new observational analyses have led to the first
astrophysical tests of gravity. We summarized current constraints,
upcoming surveys and prospects for new tests.  

\section{Acknowledgements}
We are indebted to Niayesh Afshordi, Alex Borisov, Claudia de Rham, Gia Dvali, Gregory Gabadadze, Ghazal Geshnizjani, Kurt
Hinterbichler, Stefan Hofmann, Wayne Hu, Lam Hui, Kazuya Koyama, Marcos Lima, 
Alberto Nicolis, Fabian Schmidt, Roman Scoccimarro, Ravi Sheth, Andrew
Tolley, Mark Trodden, Amol Upadhye, Jean-Philippe Uzan, Daniel Wesley
and Mark Wyman for many enlightening discussions over the years. We
thank Pengjie Zhang, Jacek Guzik and Masahiro Takada for collaborative work
on which parts of Sec.~3 are based. We are grateful to  
Fabian Schmidt for comments on the manuscript and for providing us
Figs. 3 and 6.  
This work was partially supported by funds from the University of Pennsylvania and NSERC of Canada (J.K.), and by NSF grant AST-0607667 (B.J.).

\end{document}